\documentclass[12pt,a4paper,english,amssymb,nofootinbib,superscriptaddress,showpacs]{revtex4}
\usepackage{lmodern}
\usepackage{lmodern}
\usepackage[T1]{fontenc}
\usepackage[latin9]{inputenc}
\usepackage{color}
\usepackage{babel}
\usepackage{amsmath}
\usepackage{amssymb}
\usepackage{esint}
\usepackage[unicode=true,pdfusetitle,
 bookmarks=true,bookmarksnumbered=false,bookmarksopen=false,
 breaklinks=false,pdfborder={0 0 1},backref=false,colorlinks=true]
 {hyperref}
\hypersetup{
 linkcolor=blue,citecolor=blue}

\makeatletter

\pdfpageheight\paperheight
\pdfpagewidth\paperwidth

\@ifundefined{textcolor}{}
{%
 \definecolor{BLACK}{gray}{0}
 \definecolor{WHITE}{gray}{1}
 \definecolor{RED}{rgb}{1,0,0}
 \definecolor{GREEN}{rgb}{0,1,0}
 \definecolor{BLUE}{rgb}{0,0,1}
 \definecolor{CYAN}{cmyk}{1,0,0,0}
 \definecolor{MAGENTA}{cmyk}{0,1,0,0}
 \definecolor{YELLOW}{cmyk}{0,0,1,0}
}

\usepackage{babel}
\usepackage{babel}

\usepackage{babel}

\usepackage{latexsym}\usepackage{bm}

\makeatother

\begin{document}

\title{AdS-plane wave and pp-wave solutions of generic gravity theories}

\author{Metin Gürses}

\email{gurses@fen.bilkent.edu.tr}

\affiliation{{\small{}Department of Mathematics, Faculty of Sciences}\\
 {\small{}Bilkent University, 06800 Ankara, Turkey}}

\author{Tahsin Ça\u{g}r\i{} \c{S}i\c{s}man}

\email{tahsin.c.sisman@gmail.com}

\affiliation{Centro de Estudios Cient\'{i}ficos (CECS), Casilla 1469, Valdivia,
Chile}

\affiliation{Department of Astronautical Engineering,\\
 University of Turkish Aeronautical Association, 06790 Ankara, Turkey}

\author{Bayram Tekin}

\email{btekin@metu.edu.tr}

\affiliation{Department of Physics,\\
 Middle East Technical University, 06800 Ankara, Turkey}

\date{\today}
\begin{abstract}
We construct the AdS-plane wave solutions of generic gravity theory
built on the arbitrary powers of the Riemann tensor and its derivatives
in analogy with the pp-wave solutions. In constructing the wave solutions
of the generic theory, we show that the most general two tensor built
from the Riemann tensor and its derivatives can be written in terms
of the traceless-Ricci tensor. Quadratic gravity theory plays a major
role; therefore, we revisit the wave solutions in this theory. As
examples to our general formalism, we work out \textit{\emph{the six-dimensional
conformal gravity and its nonconformal deformation as well as the
tricritical gravity, the Lanczos-Lovelock theory, and string-generated
cubic curvature theory.}}
\end{abstract}

\pacs{04.50.--h, 04.20.Jb, 04.30.--w}

\maketitle
{\footnotesize{}\tableofcontents{}}

\section{Introduction}

At short distances, Einstein's gravity is expected to be replaced
by a better behaved effective theory with more powers of curvature
and its derivatives which can be written in the most general form
(with no matter fields) as 
\begin{equation}
I=\int d^{D}x\sqrt{-g}\, f\left(g^{\alpha\beta},R_{\phantom{\mu}\nu\gamma\sigma}^{\mu},\nabla_{\rho}R_{\phantom{\mu}\nu\gamma\sigma}^{\mu},\dots,\left(\nabla_{\rho_{1}}\nabla_{\rho_{2}}\dots\nabla_{\rho_{M}}\right)R_{\phantom{\mu}\nu\gamma\sigma}^{\mu},\dots\right).\label{eq:Functional_general_theory}
\end{equation}
Although we will give solutions to this theory, often times, it is
more convenient to take the following power series version 
\begin{align}
I=\int d^{D}x\,\sqrt{-g}\,\Biggl\{ & \frac{1}{\kappa}\left(R-2\Lambda_{0}\right)\nonumber \\
 & +\alpha R^{2}+\beta R_{\mu\nu}R^{\mu\nu}+\gamma\left(R_{\mu\nu\rho\sigma}R^{\mu\nu\rho\sigma}-4R_{\mu\nu}R^{\mu\nu}+R^{2}\right)\nonumber \\
 & +\sum_{n=3}^{\infty}C_{n}\left(\text{Riem, Ric, R, }\nabla\text{Riem, }\dots\right)^{n}\Biggr\},\label{eq:Generic_gravity}
\end{align}
where we have added a bare cosmological constant $\Lambda_{0}$--not
required at short distances--but plays a major phenomenological role
at long distances. We have separated the quadratic parts as they will
play a role in the construction of solutions to the generic theory
and we have also organized the third term in the quadratic curvature
modifications into the Gauss-Bonnet form which is easier to handle
as it gives second order equations in the metric just like the Einstein's
theory. Note that the third line represents all other possible contractions
of the Riemann tensor and its derivatives which provide beyond fourth
order field equations in the metric; for example, terms such as $R\square R$
are also included in that summation. In a microscopic theory, such
as the string theory, the parameters $\alpha$, $\beta$, $\gamma$,
$C_{n}$, $\Lambda_{0}$, $\kappa$ are expected to be computed and
some of them obviously vanish due to the constraints such as unitarity,
supersymmetry, etc. Here, to stay as generic as possible, without
focusing too much on such constraints, we shall consider (\ref{eq:Functional_general_theory})
and (\ref{eq:Generic_gravity}) to be the \emph{theory} and seek exact
solutions for it. Of course we shall give some specific examples as
noted in the Abstract. It should be mentioned that not all the theories
in the form (\ref{eq:Generic_gravity}) give healthy, stable theories
when linearized about their vacua. For example, most theories yield
higher time derivative free theories that have Ostrogradsky instability
when small interactions are added. These considerations do not deter
us from studying the most general action given as (\ref{eq:Functional_general_theory})
or (\ref{eq:Generic_gravity}) since our theories include all possible
viable theories as well as the instability-plagued ones.\textit{\emph{
We know that the $f\left(R\right)$ gravity theories are free from
Ostrogradsky instability. In addition to this subclass of (\ref{eq:Generic_gravity}),
to obtain a theory that is free from Ostrogradsky instability is still
an open question.}}

Unlike the case of Einstein's gravity where books compiling exact
solutions exist \cite{Stephani,Podolsky}, there are only a few solutions
known for some variants or restricted versions of the theory (\ref{eq:Generic_gravity}),
see for example \cite{Boulware,Gueven,Amati,Horowitz-Steif,Horowitz-Tseytlin,Banks-Green,Coley,Hervik,Hassaine,fRicci,Ahmedov-Aliev,MalekThesis}.
In \cite{Gurses-PRL}, we have briefly sketched the proof that the
AdS-waves (both plane and spherical) that solve Einstein's gravity
and the quadratic gravity also solve the generic theory (\ref{eq:Generic_gravity}),
needless to say, with modified parameters. Here, we shall give a detailed
proof for AdS-plane wave case with a direct approach based on the
proof that pp-wave solutions of Einstein's gravity and the quadratic
gravity are solutions to the theory with $\Lambda_{0}=0$. As we shall
see, having a nonzero $\Lambda_{0}$ complicates the matter in a great
deal. (AdS-spherical waves require a separate attention which we shall
come back to in another work.)

In this work, we will exclusively be interested in the exact solutions
and not perturbative excitations about the maximally symmetric vacua
of the theory. Nevertheless, the fact that these exact solutions \emph{linearize}
the field equations just like the perturbative excitations, leads
to the following remarkable consequence: These metrics can be used
to test the unitarity of the underlying theory and to find the excitation
masses and the degrees of freedom of the spin-2 sector (There is an
important caveat here, if the theory for these test metrics turns
out to be nonunitary, then the theory is nonunitary. But, if the theory
turns out to be unitary for these test metrics, then this does not
mean that the theory is unitary one still has to check the unitarity
of the spin-0 sector). In the examples that we shall study here, the
procedure will be apparent.

AdS-plane waves \cite{Alishah,Gullu-Gurses} and AdS-spherical waves
\cite{gurses1} of quadratic gravity theories played a central role
in \cite{Gurses-PRL}. We shall study here the AdS-plane wave (sometimes
called the Siklos metric \cite{Siklos}) given as\textit{\emph{ 
\begin{equation}
ds^{2}=\frac{\ell^{2}}{z^{2}}\left(2dudv+d\vec{x}\cdot d\vec{x}+dz^{2}\right)+2V\left(u,\vec{x},z\right){\rm d}u^{2},\label{eq:AdS-plane_metric}
\end{equation}
}}where $u$ and $v$ are null coordinates,\textit{\emph{ $\vec{x}=\left(x^{i}\right)$
with $i=1,\dots,D-3$}}, and $\ell$ is the AdS radius related to
the effective cosmological constant as $\Lambda=-\frac{\left(D-1\right)\left(D-2\right)}{2\ell^{2}}$.
For this $D$-dimensional metric, the Ricci tensor can be computed
to be 
\begin{equation}
R_{\mu\nu}=-\frac{\left(D-1\right)}{\ell^{2}}g_{\mu\nu}+\rho\lambda_{\mu}\lambda_{\nu},
\end{equation}
where the vector is $\lambda_{\mu}=\delta_{\mu}^{u}$ and the scalar
function is\textit{\emph{ 
\begin{equation}
\rho\equiv-\left(\square+\frac{4z}{\ell^{2}}\partial_{z}-\frac{2\left(D-3\right)}{\ell^{2}}\right)V,
\end{equation}
with $\square\equiv\nabla^{\mu}\nabla_{\mu}$ and $\nabla_{\mu}$
is compatible with the full metric (\ref{eq:AdS-plane_metric}). For
these spacetimes, as we showed \cite{Gurses-PRL}, the field equations
of (\ref{eq:Generic_gravity}) reduce to 
\begin{equation}
eg_{\mu\nu}+a_{0}S_{\mu\nu}+a_{1}\square S_{\mu\nu}+\dots+a_{n}\square^{n}S_{\mu\nu}+\dots=0,\label{eq:Full_theory_EoM_sum}
\end{equation}
where $S_{\mu\nu}$ is the traceless-Ricci tensor. Taking the trace
gives $e=0$, which determines the effective cosmological constant
of the theory. $S_{\mu\nu}=0$, which is the Einsteinian solution,
naturally solves the full theory. For (\ref{eq:AdS-plane_metric})
to be a solution to cosmological Einstein's theory, $V$ satisfies
the $\rho=0$ equation, namely 
\begin{equation}
\left(\square+\frac{4z}{\ell^{2}}\partial_{z}-\frac{2\left(D-3\right)}{\ell^{2}}\right)V\left(u,\vec{x},z\right)=0,
\end{equation}
whose solution is}}%
\footnote{\textit{\emph{Since the equation is linear in $V$, the most general
solution will be a sum or an integral over the arbitrary parameter
$\xi$ if no further condition is given. As the most general solution
is easy to write we do not depict it here.}}%
} 
\begin{equation}
V\left(u,\vec{x},z\right)=z^{\frac{D-5}{2}}\left[c_{1}I_{\frac{D-1}{2}}\left(z\xi\right)+c_{2}K_{\frac{D-1}{2}}\left(z\xi\right)\right]\sin(\vec{\xi}\cdot\vec{x}+c_{3}),\label{eq:Einstein_gen_soln}
\end{equation}
\textit{\emph{with $I$, $K$ being the modified Bessel functions,
$\left|\vec{\xi}\right|=\xi$ and $c_{i}$'s are arbitrary functions
of the null coordinate $u$ \cite{Chamblin,Gullu-Gurses}. Further
assuming $\xi=0$, the solution becomes \cite{Kaigorodov} (see also
\cite{Chamblin}) 
\begin{equation}
V\left(u,z\right)=c\left(u\right)z^{D-3},\label{eq:Kaigorodov}
\end{equation}
where we omit the other solution, that is $\frac{1}{z^{2}}$, since
it can be added to the ``background'' AdS part which is the $V=0$
case of the metric (\ref{eq:AdS-plane_metric}). This is all in cosmological
Einstein's theory. But, observe that neither (\ref{eq:Einstein_gen_soln})
or (\ref{eq:Kaigorodov}) depend explicitly on the cosmological constant
of the theory. The dependence of the metric on the cosmological constant
is only in the ``AdS background'' part. This leads to the fact that
these Einsteinian solutions remain intact in the most general theory
(\ref{eq:Generic_gravity}) with the only adjustment that the cosmological
constant that appears in the AdS background part depends on the parameters
of the the full theory. As we shall show in Sec.~\ref{sec:Einstein_soln_gen_theo},
one can find the cosmological constant, which will be determined by
nonderivative terms in the action (\ref{eq:Generic_gravity}), without
going through the cumbersome task of finding the field equations.}}

\textit{\emph{Now, let us consider the same metric as a solution to
quadratic gravity. AdS-plane wave solutions of quadratic gravity again
solve the field equations of the full theory (\ref{eq:Full_theory_EoM_sum})
as it will be more apparent when the field equations are represented
in the factorized form (\ref{eq:Multiplicative_EoM_Smn}). In this
case, the metric function $V$ satisfies a more complicated fourth
order equation}} 
\begin{equation}
\left(\square+\frac{4z}{\ell^{2}}\partial_{z}-\frac{2\left(D-3\right)}{\ell^{2}}-M^{2}\right)\left(\square+\frac{4z}{\ell^{2}}\partial_{z}-\frac{2\left(D-3\right)}{\ell^{2}}\right)V\left(u,\vec{x},z\right)=0,\label{eq:Quadratic_eqn}
\end{equation}
where the ``mass'' parameter reads 
\begin{equation}
M^{2}\equiv-\frac{1}{\beta}\left(\frac{1}{\kappa}-\frac{2}{\ell^{2}}\left(\left(D-1\right)\left(D\alpha+\beta\right)+\left(D-3\right)\left(D-4\right)\gamma\right)\right).\label{eq:M2}
\end{equation}
Assuming $M^{2}\ne0$, that is the nondegenerate case, the most general
solution of (\ref{eq:Quadratic_eqn}) can be constructed from two
second order parts; one the pure Einstein's theory 
\begin{equation}
\left(\square+\frac{4z}{\ell^{2}}\partial_{z}-\frac{2\left(D-3\right)}{\ell^{2}}\right)V_{a}\left(u,\vec{x},z\right)=0,
\end{equation}
and the other ``massive'' version of the theory 
\begin{equation}
\left(\square+\frac{4z}{\ell^{2}}\partial_{z}-\frac{2\left(D-3\right)}{\ell^{2}}-M^{2}\right)V_{b}\left(u,\vec{x},z\right)=0,
\end{equation}
with $V=V_{a}+V_{b}$. Since we already know $V_{a}$ from (\ref{eq:Einstein_gen_soln}),
let us write $V_{b}$ 
\begin{equation}
V_{b}\left(u,\vec{x},z\right)=z^{\frac{D-5}{2}}\left[c_{b,1}I_{\nu_{b}}\left(z\xi_{b}\right)+c_{b,2}K_{\nu_{b}}\left(z\xi_{b}\right)\right]\sin(\vec{\xi}_{b}\cdot\vec{x}+c_{b,3}),
\end{equation}
where $\nu_{b}=\frac{1}{2}\sqrt{\left(D-1\right)^{2}+4\ell^{2}M^{2}}$
\cite{Gullu-Gurses}. If, on the other hand, $M^{2}=0$, which also
includes the critical gravity \cite{LuPope,DeserLiu}, the solution
becomes highly complicated in the most general case $\xi\ne0$ (it
was given in the Appendix of \cite{Gullu-Gurses} which we do not
reproduce here). For the special case of $\xi=0$, the solution is%
\footnote{When $\nu_{b}=0$, Breitenlohner-Freedman (BF) bound \cite{BF} is
saturated and the solution turns into a logarithmic one given in \cite{Gullu-Gurses}
as 
\[
V\left(u,z\right)=c_{1}z^{D-3}+z^{\frac{D-5}{2}}\left[c_{1}+c_{2}\ln\left(\frac{z}{\ell}\right)\right].
\]
} 
\begin{equation}
V\left(u,z\right)=c_{a,1}z^{D-3}+z^{\frac{D-5}{2}}\left(c_{b,1}z^{\left|\nu_{b}\right|}+c_{b,2}z^{-\left|\nu_{b}\right|}\right).
\end{equation}
for $M^{2}\ne0$, and 
\begin{equation}
V\left(u,z\right)=c_{1}z^{D-3}+\frac{1}{D-1}\left(c_{2}z^{D-3}-\frac{c_{3}}{z^{2}}\right)\ln\left(\frac{z}{\ell}\right),
\end{equation}
for $M^{2}=0$. Note that all $c_{a,i}$'s and $c_{b,i}$'s appearing
in the solutions of quadratic gravity are arbitrary functions of $u$.

It was announced in \cite{Gurses-PRL} that these AdS-plane wave solutions
of Einstein's gravity and the quadratic gravity also solve the most
general theory defined by the action (\ref{eq:Generic_gravity}) with
redefined parameters that are $M^{2}$ and $\ell^{2}$. This work
expounds upon the results of \cite{Gurses-PRL}. In doing this, we
show that the pp-wave spacetimes in the Kerr-Schild form having the
metric 
\begin{equation}
ds^{2}=2dudv+d\vec{x}\cdot d\vec{x}+2V\left(u,\vec{x}\right)du^{2},\label{eq:pp-wave_KS_intro}
\end{equation}
where $\vec{x}=\left(x^{i}\right)$ with $i=1,\dots,D-2$, and the
AdS-plane wave spacetimes have analogous algebraic properties, and
with these specific properties in both cases the highly complicated
field equations of generic gravity theory reduce to somewhat simpler
equations that admit exact solutions as exemplified above. For the
pp-wave spacetimes (\ref{eq:pp-wave_KS_intro}), in complete analogy
with (\ref{eq:Full_theory_EoM_sum}), the field equations for the
full theory (\ref{eq:Generic_gravity}) reduce to\textit{\emph{ 
\begin{equation}
a_{0}R_{\mu\nu}+a_{1}\square R_{\mu\nu}+\dots+a_{n}\square^{n}R_{\mu\nu}+\dots=0.\label{eq:Full_theory_EoM_sum-pp-wave}
\end{equation}
which is solved by the Einsteinian solution $R_{\mu\nu}=0$. Once
one considers plane waves, which is a subclass of pp-wave spacetimes
with the metric 
\begin{equation}
ds^{2}=2dudv+d\vec{x}\cdot d\vec{x}+h_{ij}\left(u\right)x^{i}x^{j}{\rm d}u^{2},\label{eq:Plane_wave}
\end{equation}
where $\vec{x}=\left(x^{i}\right)$ with $i=1,\dots,D-2$, and $h_{ij}$
is symmetric and traceless, $R_{\mu\nu}$ vanishes and one has a solution
of (\ref{eq:Full_theory_EoM_sum-pp-wave}) for any $h_{ij}$. Thus,
the plane-wave solutions of Einstein's gravity solve the generic theory
}}\cite{Gueven}\textit{\emph{. The pp-wave metric (\ref{eq:pp-wave_KS_intro})
solves Einstein's gravity if the metric function $V$ satisfies the
Laplace equation for the $\left(D-2\right)$-dimensional space, and
the fact that these solutions solve the generic gravity }}theory \textit{\emph{(\ref{eq:Full_theory_EoM_sum-pp-wave})
was first shown in \cite{Horowitz-Steif}. In addition, if vanishing
scalar invariant spacetimes, of which (\ref{eq:pp-wave_KS_intro})
is a member, satisfy $\square R_{\mu\nu}=0$, the field equations
of (\ref{eq:Full_theory_EoM_sum-pp-wave}) again reduce to the Einsteinian
ones \cite{Coley}. In addition to these Einstein's gravity-based
considerations, as we shall show below by putting (\ref{eq:Full_theory_EoM_sum-pp-wave})
in the factorized form (\ref{eq:Multiplicative_EoM_Rmn}), one can
observe that the pp-wave solutions of quadratic curvature gravity
which satisfy $\left(b_{1}\square+b_{0}\right)R_{\mu\nu}=0$ also
solve the generic theory. Note that one can extend these solutions
to theories with}}\textit{\textcolor{red}{\emph{ }}}\textit{\emph{pure
radiation sources, that is $T_{\mu\nu}dx^{\mu}dx^{\nu}=T_{uu}du^{2}$.
With this kind of sources and metrics satisfying $\square R_{\mu\nu}=0$,
the field equations take the form $a_{0}R_{uu}=T_{uu}$, and the case
of $T_{uu}=T_{uu}\left(u\right)$ was considered in \cite{Gueven,Horowitz-Steif,Coley}.
A solution to $a_{0}R_{uu}=T_{uu}\left(u\right)$ can be found, for
example, by relaxing the traceless condition on $h_{ij}\left(u\right)$
of (\ref{eq:Plane_wave}), then one has simply the algebraic equation
$a_{0}\sum_{i=1}^{D-2}h_{ii}\left(u\right)=-T_{uu}\left(u\right)$
\cite{Gueven}.}}

The layout of the paper is as follows: In Section II, pp-wave spacetimes
in generic gravity theory are discussed to set the stage for AdS-plane
waves discussed in Section III which also includes the proof of the
theorem that a generic two tensor can be reduced to \textit{\emph{a
linear combination of $g_{\mu\nu}$, $S_{\mu\nu}$, and higher orders
of $S_{\mu\nu}$ (such as, for example, $\square^{n}\, S_{\mu\nu}$).
Section IV is devoted to the field equations of quadratic gravity
for pp-wave and AdS-wave ansatze which play a major role in generic
gravity theories. In Section V, we study the wave solutions of $f\left(R_{\alpha\beta}^{\mu\nu}\right)$
theories where the action depends on the Riemann tensor but not on
its derivatives. As two examples, we study the cubic gravity generated
by string theory and the Lanczos-Lovelock theory. In Section VI, we
show that Einsteinian wave solutions solve the generic gravity theory
and as an example, we study the AdS-plane wave solutions of the six-dimensional
conformal gravity and its nonconformal deformation as well as the
tricritical gravity. In the Appendices, we expound upon some of the
calculations given in the text.}}

\section{pp-Wave Spacetimes in Generic Gravity Theory}

As discussed above, analogies with the pp-wave solution will play
a role in our proof so we first study the simpler pp-wave case. The
pp-wave spacetime is a spacetime with plane-fronted parallel rays
(for further properties of pp-waves see, for example, \cite{Blau,Classification}).
A subclass of these metrics can be put into the Kerr-Schild form as
\begin{equation}
g_{\mu\nu}=\eta_{\mu\nu}+2V\lambda_{\mu}\lambda_{\nu},\label{eq:pp-wave_KS}
\end{equation}
where $\eta_{\mu\nu}$ is the Minkowski metric and the following relations
hold 
\begin{equation}
\lambda^{\mu}\lambda_{\mu}=0,\qquad\nabla_{\mu}\lambda_{\nu}=0,\qquad\lambda^{\mu}\partial_{\mu}V=0.
\end{equation}
The pp-wave spacetimes have special algebraic properties. The Riemann
and Ricci tensors of pp-waves in the Kerr-Schild form are classified
as Type N according to the ``null alignment classification'' \cite{NullAlignment,BoostWeight}.
When the Riemann and Ricci tensors are calculated by using (\ref{eq:pp-wave_KS}),
they, respectively, become 
\begin{equation}
R_{\mu\alpha\nu\beta}=\lambda_{\mu}\lambda_{\beta}\partial_{\alpha}\partial_{\nu}V+\lambda_{\alpha}\lambda_{\nu}\partial_{\mu}\partial_{\beta}V-\lambda_{\mu}\lambda_{\nu}\partial_{\alpha}\partial_{\beta}V-\lambda_{\alpha}\lambda_{\beta}\partial_{\mu}\partial_{\nu}V,\label{eq:Riemann_pp-wave}
\end{equation}
and 
\begin{equation}
R_{\mu\nu}=-\lambda_{\mu}\lambda_{\nu}\partial^{2}V,\label{eq:Ricci_pp-wave}
\end{equation}
which make the Type-N properties explicit. With these forms of the
Riemann and Ricci tensors, notice that any contraction with the $\lambda^{\mu}$
vector yields zero. The scalar curvature is zero for the metric (\ref{eq:pp-wave_KS}).
Besides the scalar curvature, it has vanishing scalar invariants (VSI).
Since the Riemann and Ricci tensors are of Type N, and the scalar
curvature is zero, the pp-wave spacetimes are also Type-N Weyl. Lastly,
since the $\lambda^{\mu}$ vector is covariantly constant, it is nonexpanding,
shear-free, and nontwisting; therefore, the pp-wave metrics belong
to the Kundt class of metrics.

The two tensors of pp-wave spacetimes also have a special structure:
\textit{\emph{Any second rank tensor constructed from the Riemann
tensor and its covariant derivatives can be written as a linear combination
of $R_{\mu\nu}$ and higher orders of $R_{\mu\nu}$ (such as, for
example, $\square^{n}\, R_{\mu\nu}$ with $n$ a positive integer).
This result follows from the corresponding property of Type-N Weyl,
Type-N Ricci spacetimes given in \cite{Hervik} as the pp-wave spacetimes
in the Kerr-Schild form share these properties. Although the pp-wave
result is implied in \cite{Hervik}, here we provide the proof along
the lines of \cite{Horowitz-Steif} since it gives some insight on
the corresponding proof for AdS-plane wave given below.}}

\subsection{Two-tensors in pp-wave spacetime}

A generic two tensor of the pp-wave spacetimes can be, symbolically,
represented as 
\begin{equation}
\left[R^{n_{0}}\left(\nabla^{n_{1}}R\right)\left(\nabla^{n_{2}}R\right)\dots\left(\nabla^{n_{m}}R\right)\right]_{\mu\nu},\label{eq:Generic_two-tensor_pp-wave}
\end{equation}
where $R$ denotes the Riemann tensor, $\nabla^{n_{i}}R$ represents
the $\left(0,n_{i}+4\right)$ rank tensor constructed by $n_{i}$
number of covariant derivatives acting on the Riemann tensor, so the
term in $\left[\dots\right]_{\mu\nu}$ is a $\left(0,4n_{0}+4m+\sum_{i=1}^{m}n_{i}\right)$
rank tensor whose indices are contracted until two indices, $\mu$
and $\nu$, are left free. Here, the important point to notice is
that each Riemann tensor has two $\lambda$'s (\ref{eq:Riemann_pp-wave}),
so in total there are $2\left(n_{0}+m\right)$ number of $\lambda$
vectors. The remaining tensor structure involves just $\nabla^{n}V$'s.

Here is what we will prove: \emph{The generic two tensors of the form
(\ref{eq:Generic_two-tensor_pp-wave}) will boil down to a linear
combination of $R_{\mu\nu}$ and }\textit{\emph{$\square^{n}\, R_{\mu\nu}$'}}\textit{s}\textit{\emph{.}}

The first step of the proof is showing that the $\lambda$ vector
cannot make a nonzero contraction. It is easy to show this by using
mathematical induction. With the identity $\lambda^{\mu}\partial_{\mu}V=0$,
the $\lambda$ contraction of the term $\nabla^{2}V$ is simply zero
\begin{equation}
\lambda^{\mu}\nabla_{\nu}\partial_{\mu}V=0,
\end{equation}
after using the fact that $\lambda$ is covariantly constant. Then,
to show that the $\lambda$ contraction of the term $\nabla^{n}V$
reduces to a lower order term, first observe that 
\begin{equation}
\lambda^{\mu_{j}}\nabla_{\mu_{1}}\dots\nabla_{\mu_{j}}\dots\nabla_{\mu_{n}}V=\nabla_{\mu_{1}}\left(\lambda^{\mu_{j}}\nabla_{\mu_{2}}\dots\nabla_{\mu_{j}}\dots\nabla_{\mu_{n}}V\right).
\end{equation}
Secondly, when $\lambda$ is contracted with the first covariant derivative,
by using $\left[\nabla_{\alpha},\nabla_{\beta}\right]V^{\rho}=R_{\alpha\beta\phantom{\rho}\sigma}^{\phantom{\alpha\beta}\rho}V^{\sigma}$
and $\lambda^{\mu}R_{\mu\alpha\nu\beta}=0$, one has 
\begin{align}
\lambda^{\mu_{1}}\nabla_{\mu_{1}}\nabla_{\mu_{2}}\dots\nabla_{\mu_{n}}V & =\lambda^{\mu_{1}}\left[\nabla_{\mu_{1}},\nabla_{\mu_{2}}\right]\dots\nabla_{\mu_{n}}V+\lambda^{\mu_{1}}\nabla_{\mu_{2}}\nabla_{\mu_{1}}\dots\nabla_{\mu_{n}}V\nonumber \\
 & =\lambda^{\mu_{1}}\nabla_{\mu_{2}}\nabla_{\mu_{1}}\dots\nabla_{\mu_{n}}V,
\end{align}
which completes the reduction of the $n^{{\rm th}}$ order term to
the $\left(n-1\right)^{{\rm th}}$ order. Thus, $\lambda$ cannot
make a nonzero contraction either with other $\lambda$'s or with
$\nabla^{n}V$'s.

Although we achieved our goal, let us discuss another proof of this
step which gives an insight for the corresponding discussion in the
AdS-plane wave case. For the pp-wave metrics in the Kerr-Schild form,
one can choose the coordinates in such a way that the metric takes
the form 
\begin{equation}
ds^{2}=2dudv+d\vec{x}\cdot d\vec{x}+2V\left(u,\vec{x}\right)du^{2},\label{eq:pp-wave_coordinate_choice}
\end{equation}
where $\vec{x}=\left(x^{i}\right)$ with $i=1,\dots,D-2$, and $u$
and $v$ are null coordinates, so 
\begin{equation}
\lambda_{\mu}dx^{\mu}=du\Rightarrow\lambda^{\mu}\partial_{\mu}=\partial_{v}\Rightarrow\lambda^{\mu}\partial_{\mu}V=\partial_{v}V=0.
\end{equation}
With this choice of the metric, $\nabla_{\mu}\lambda^{\nu}=0$ leads
$\Gamma_{\mu\sigma}^{\nu}\lambda^{\sigma}=0$. Then, let us look at
the expansion of $\nabla^{n}V$ which has the form 
\begin{align}
\nabla_{\mu_{1}}\nabla_{\mu_{2}}\dots\nabla_{\mu_{n}}V= & \partial_{\mu_{1}}\partial_{\mu_{2}}\dots\partial_{\mu_{n}}V-\left(\partial_{\mu_{1}}\partial_{\mu_{2}}\dots\partial_{\mu_{n-2}}\Gamma_{\mu_{n-1}\mu_{n}}^{\sigma_{1}}\right)\partial_{\sigma_{1}}V\nonumber \\
 & -\Gamma_{\mu_{n-1}\mu_{n}}^{\sigma_{1}}\partial_{\mu_{1}}\partial_{\mu_{2}}\dots\partial_{\mu_{n-2}}\partial_{\sigma_{1}}V\nonumber \\
 & -\dots-\left(-1\right)^{n-1}\Gamma_{\mu_{1}\mu_{2}}^{\sigma_{1}}\Gamma_{\sigma_{1}\mu_{3}}^{\sigma_{2}}\dots\Gamma_{\sigma_{n-2}\mu_{n}}^{\sigma_{n-1}}\partial_{\sigma_{n-1}}V.
\end{align}
The structures appearing in this expansion are the Christoffel connection,
partial derivatives of both $V$ and the Christoffel connection. When
one has a $\lambda$ contraction, some terms involve contraction of
$\lambda$ with one of the partial derivatives acting on $V$ which
yields an immediate zero since $\lambda^{\mu}=\delta_{v}^{\mu}$ and
$\partial_{v}V=0$. In addition, a $\lambda$ contraction with a Christoffel
connection also yields zero. On the other hand, if $\lambda$ is contracted
with one of the partial derivatives acting on a Christoffel connection,
one needs to use the definition of the Riemann tensor, for example
as 
\begin{align}
\lambda^{\mu_{j}}\partial_{\mu_{1}}\dots\partial_{\mu_{j}}\dots\partial_{\mu_{n-2}}\Gamma_{\mu_{n-1}\mu_{n}}^{\sigma_{1}}= & \partial_{\mu_{1}}\dots\partial_{\mu_{n-2}}\left(\lambda^{\mu_{j}}\partial_{\mu_{j}}\Gamma_{\mu_{n-1}\mu_{n}}^{\sigma_{1}}\right)\nonumber \\
= & \partial_{\mu_{1}}\dots\partial_{\mu_{n-2}}\left[\lambda^{\mu_{j}}\left(R_{\phantom{\sigma_{1}}\mu_{n}\mu_{j}\mu_{n-1}}^{\sigma_{1}}+\partial_{\mu_{n-1}}\Gamma_{\mu_{j}\mu_{n}}^{\sigma_{1}}\right.\right.\nonumber \\
 & \phantom{\partial_{\mu_{1}}\dots\partial_{\mu_{n-2}}\left[\lambda^{\mu_{j}}\right.}\left.\left.-\Gamma_{\mu_{j}\alpha}^{\sigma_{1}}\Gamma_{\mu_{n-1}\mu_{n}}^{\alpha}+\Gamma_{\mu_{n-1}\alpha}^{\sigma_{1}}\Gamma_{\mu_{j}\mu_{n}}^{\alpha}\right)\right],
\end{align}
where the terms in the square bracket are just zero since $\lambda^{\mu}R_{\mu\alpha\nu\beta}=0$
and $\Gamma_{\mu\sigma}^{\nu}\lambda^{\sigma}=0$.

Since $\lambda$ cannot make a nonzero contraction, there should be
at most two $\lambda$'s, that is one Riemann tensor, so the nonzero
terms of the form (\ref{eq:Generic_two-tensor_pp-wave}) reduce to
\begin{equation}
R_{\mu\nu},\qquad{\rm or}\qquad\left[\nabla^{2n}R\right]_{\mu\nu},
\end{equation}
where even number of covariant derivatives is required to have a two
tensor. After determining the nonzero terms required by the first
step of the proof, now in the second step, let us discuss the structure
of these nonzero terms of the form $\left[\nabla^{n}R\right]_{\mu\nu}$.
In obtaining a two tensor by contracting the indices of $\left[\nabla^{n}R\right]_{\mu\nu}$,
one should either have 
\begin{equation}
g^{\alpha\beta}\nabla_{\mu_{1}}\nabla_{\mu_{2}}\dots\nabla_{\mu_{2n}}R_{\mu\alpha\nu\beta}=\nabla_{\mu_{1}}\nabla_{\mu_{2}}\dots\nabla_{\mu_{2n}}R_{\mu\nu},\label{eq:First_nonzero_pp}
\end{equation}
or 
\begin{equation}
\nabla_{\mu_{1}}\nabla_{\mu_{2}}\dots\nabla^{\alpha}\dots\nabla^{\beta}\dots\nabla_{\mu_{2n-2}}R_{\mu\alpha\nu\beta}.\label{eq:Second_nonzero_pp_ini}
\end{equation}
In (\ref{eq:Second_nonzero_pp_ini}), one can rearrange the order
of the derivatives. Each change of order introduces a Riemann tensor,
and as we just showed, a two tensor contraction in the presence of
this additional Riemann tensor gives zero. The only nonzero part is
the original term which in the final form reads 
\begin{equation}
\nabla_{\mu_{1}}\nabla_{\mu_{2}}\dots\nabla_{\mu_{2n-2}}\nabla^{\alpha}\nabla^{\beta}R_{\mu\alpha\nu\beta}=\nabla_{\mu_{1}}\nabla_{\mu_{2}}\dots\nabla_{\mu_{2n-2}}\square R_{\mu\nu},\label{eq:Second_nonzero_pp}
\end{equation}
where we used the Bianchi identity on the Riemann tensor. Further
contractions in (\ref{eq:First_nonzero_pp}) and (\ref{eq:Second_nonzero_pp})
should be between the indices of the derivatives and as we showed
we can change the order of the derivatives without introducing an
additional term, then one has 
\[
\left[\nabla^{2n}R\right]_{\mu\nu}=\square^{n}R_{\mu\nu}.
\]
As a result, the nonzero terms are in the form $R_{\mu\nu}$ and $\square^{n}R_{\mu\nu}$,
where $n$ is a positive integer. Any two tensor of the pp-wave spacetimes
in the Kerr-Schild form is a linear combination of these terms. This
completes the proof.

Before proceeding to the field equations, note that with this result
about the two-tensors, the VSI property of the pp-waves in the Kerr-Schild
form is explicit since $R_{\mu\nu}$ is traceless.%
\footnote{For the proof of the VSI property of plane waves, see \cite{Deser}.%
}

\subsection{Field equations of the generic theory for pp-wave spacetime}

Once the above result is used, the field equations of the most general
theory (\ref{eq:Generic_gravity}) with $\Lambda_{0}=0$ reduces to
\begin{equation}
\sum_{n=0}^{N}a_{n}\square^{n}R_{\mu\nu}=0,\label{eq:EoM_gen_grav_as_sum-pp-wave}
\end{equation}
where $a_{n}$'s are constants depending on the parameters of the
theory, namely on $\kappa$, $\alpha$, $\beta$, $\gamma$, $C_{n}$,
and $N$ can be as large as possible. Note that a pp-wave metric (\ref{eq:pp-wave_KS})
solving $R_{\mu\nu}=0$ is a solution of (\ref{eq:EoM_gen_grav_as_sum-pp-wave}).
This fact was demonstrated in \cite{Horowitz-Steif} without finding
the explicit form of (\ref{eq:EoM_gen_grav_as_sum-pp-wave}) by taking
$R_{\mu\nu}=0$ as an assumption from the beginning. The plane waves,
which are special pp-waves with $V\left(u,\vec{x}\right)=h_{ij}\left(u\right)x^{i}x^{j}$
where $h_{ij}$ is symmetric and traceless, provide a solution to
(\ref{eq:EoM_gen_grav_as_sum-pp-wave}) for any $h_{ij}$ by satisfying
$R_{\mu\nu}=0$ \cite{Gueven}. As discussed in \cite{Coley}, one
can also follow the way of constraining pp-wave spacetimes such that
$R_{\mu\nu}$ is the only nonzero two-tensor, which effectively means
$\square R_{\mu\nu}=0$, then the field equations of the generic gravity
theory reduce to the Einstein's gravity ones. On the other hand, obtaining
(\ref{eq:EoM_gen_grav_as_sum-pp-wave}) makes one to realize that
the pp-wave solutions of the quadratic gravity theory also solve the
generic gravity theory (\ref{eq:Generic_gravity}). To show this,
first notice that one can factorize (\ref{eq:EoM_gen_grav_as_sum-pp-wave})
as 
\begin{equation}
\prod_{n=1}^{N}\left(\square+b_{n}\right)R_{\mu\nu}=0,\label{eq:Multiplicative_EoM_Rmn}
\end{equation}
where $b_{n}$'s are constants depending on the original parameters
of the theory. Here, $b_{n}$'s can be real or complex and once they
are complex, they must appear in complex conjugate pairs.

To further reduce (\ref{eq:Multiplicative_EoM_Rmn}), first using
the covariantly constant property of $\lambda^{\mu}$, one has 
\begin{equation}
\square R_{\mu\nu}=\square\left(-\lambda_{\mu}\lambda_{\nu}\partial^{2}V\right)=-\lambda_{\mu}\lambda_{\nu}\square\partial^{2}V.
\end{equation}
Here, note that for \emph{any} scalar function $\phi$ (not necessarily
$V$) satisfying $\lambda^{\mu}\partial_{\mu}\phi=\partial_{v}\phi=0$,
one has 
\begin{equation}
\square\phi=g^{\mu\nu}\nabla_{\mu}\nabla_{\nu}\phi=\eta^{\mu\nu}\partial_{\mu}\partial_{\nu}\phi-\eta^{\mu\nu}\Gamma_{\mu\nu}^{\sigma}\partial_{\sigma}\phi,
\end{equation}
after also using $\Gamma_{\mu\sigma}^{\nu}\lambda^{\sigma}=0$ which
is valid in the coordinates we have chosen (\ref{eq:pp-wave_coordinate_choice}).
Here, $\eta_{\mu\nu}$ is the flat metric in null coordinates. In
addition, for (\ref{eq:pp-wave_coordinate_choice}), the Christoffel
connection has the form 
\begin{equation}
\Gamma_{\mu\nu}^{\sigma}=\lambda^{\sigma}\lambda_{\nu}\partial_{\mu}V+\lambda^{\sigma}\lambda_{\mu}\partial_{\nu}V-\lambda_{\mu}\lambda_{\nu}\eta^{\sigma\beta}\partial_{\beta}V,
\end{equation}
which leads to $\eta^{\mu\nu}\Gamma_{\mu\nu}^{\sigma}=0$; therefore,
one has 
\begin{equation}
\square\phi=\partial^{2}\phi.
\end{equation}
Furthermore, since $\partial^{2}=2\frac{\partial^{2}}{\partial u\partial v}+\hat{\partial}^{2}$,
where $\hat{\partial}^{2}\equiv\sum_{i=1}^{D-2}\frac{\partial^{2}}{\partial x^{i}\partial x^{i}}$,
and $\partial_{v}\phi=0$, we have 
\begin{equation}
\square\phi=\hat{\partial}^{2}\phi.
\end{equation}
With this property and $\partial_{v}\hat{\partial}\dots\hat{\partial}V=0$,
one has 
\begin{equation}
\square^{n}V=\hat{\partial}^{2n}V,
\end{equation}
which reduces (\ref{eq:Multiplicative_EoM_Rmn}) to 
\begin{equation}
\lambda_{\mu}\lambda_{\nu}\hat{\partial}^{2}\prod_{n=1}^{N}\left(\hat{\partial}^{2}+b_{n}\right)V=0.\label{eq:Final_EoM_pp-wave}
\end{equation}
Note that this equation is linear in $V$, so one can make an important
observation for pp-wave metrics in the Kerr-Schild form. One can consider
the pp-wave metric (\ref{eq:pp-wave_KS}) as $g_{\mu\nu}=\eta_{\mu\nu}+h_{\mu\nu}$
where $h_{\mu\nu}\equiv2V\lambda_{\mu}\lambda_{\nu}$, and with this
definition the Ricci tensor becomes 
\begin{equation}
R_{\mu\nu}=-\frac{1}{2}\partial^{2}h_{\mu\nu},
\end{equation}
after using the fact that $\lambda_{\mu}$ is covariantly constant.
Then, once one considers this form of the Ricci tensor and $\square^{n}R_{\mu\nu}=\partial^{2n}R_{\mu\nu}$
in either (\ref{eq:EoM_gen_grav_as_sum-pp-wave}) or (\ref{eq:Multiplicative_EoM_Rmn}),
it is clear that the field equations of the generic theory (\ref{eq:Generic_gravity})
for pp-waves are linear in $h_{\mu\nu}$ as in the case of a perturbative
expansion of the field equations around a flat background for a small
metric perturbation $\left\Vert h\right\Vert \equiv\left\Vert g-\eta\right\Vert \ll1$.

This observation suggests that there are two possible ways to find
the field equations of the generic gravity theory for pp-waves: 1-
Deriving the field equations and directly putting the pp-wave metric
ansatz (\ref{eq:pp-wave_KS}) in them; or 2- Linearizing the derived
field equations around the flat background and putting $h_{\mu\nu}=2V\lambda_{\mu}\lambda_{\nu}$
in these linearized equations. Although the second way involves an
additional linearization step, the idea itself provides a shortcut
in finding the field equations of pp-waves for a gravity theory described
with a Lagrangian density which is constructed by the Riemann tensor
but not its derivatives. Namely, due to linearization in the field
equations, only up to the quadratic curvature order of these theories
contributes to the field equations. This idea is made explicit in
the examples discussed in Sec.~\ref{sec:fRiem_solns}. Lastly, since
$h_{\mu\nu}=2V\lambda_{\mu}\lambda_{\nu}$ is transverse, $\partial_{\mu}h^{\mu\nu}=0$,
and traceless, $\eta^{\mu\nu}h_{\mu\nu}=0$, to find the field equations
by following the second way, one needs only the linearized field equations
for the transverse-traceless metric perturbation.

Assuming nonvanishing and distinct $b_{n}$'s, the most general solution
of (\ref{eq:Final_EoM_pp-wave}) is 
\begin{equation}
V=V_{E}+\Re\left(\sum_{n=1}^{N}V_{n}\right),
\end{equation}
where $V_{E}$ is the solution to Einstein's theory, namely $\hat{\partial}^{2}V_{E}=0$,
$\Re$ represents the real part, and $V_{n}$'s solve the equation
of the quadratic gravity theory, i.e. $\left(\hat{\partial}^{2}+b_{n}\right)V_{n}=0$
(lest the reader have any doubt about this equation being the quadratic
gravity theory's equation for the pp-wave, we shall show this explicitly
below). Then, the pp-wave solution of Einstein's gravity also solves
a generic gravity theory which was already known in the literature
\cite{Horowitz-Steif}. Here, the novel result is that the pp-wave
solutions of the quadratic gravity theory also solve the generic theory.
These solutions are of the form 
\begin{equation}
V_{n}\left(u,\vec{x}\right)=c_{1,n}\left(u\right)\sin(\vec{\xi}_{n}\cdot\vec{x}+c_{2,n}\left(u\right)),
\end{equation}
with $\left|\vec{\xi}_{n}\right|^{2}\equiv b_{n}$. Here, we consider
the case with real $b_{n}$ since $b_{n}$'s are related to the masses
of the perturbative excitations around flat background as $M_{n,{\rm flat}}^{2}=-b_{n}$.
What we have learned in the pp-wave case will be applied to the $\Lambda_{0}$
case below.

\section{AdS-Plane Wave Spacetimes in Generic Gravity Theory}

AdS-plane waves are a member of Kerr-Schild--Kundt metrics given as
\begin{equation}
g_{\mu\nu}=\bar{g}_{\mu\nu}+2V\lambda_{\mu}\lambda_{\nu},\label{eq:AdS-wave_KS}
\end{equation}
where $\bar{g}_{\mu\nu}$ is the AdS metric and the following relations
hold 
\begin{equation}
\lambda^{\mu}\lambda_{\mu}=0,\qquad\nabla_{\mu}\lambda_{\nu}=\xi_{(\mu}\lambda_{\nu)},\qquad\xi_{\mu}\lambda^{\mu}=0,\qquad\lambda^{\mu}\partial_{\mu}V=0.\label{eq:AdS-wave_prop}
\end{equation}
The second identity serves as a definition of $\xi$ vector where
the symmetrization convention is $\xi_{(\mu}\lambda_{\nu)}\equiv\frac{1}{2}\left(\xi_{\mu}\lambda_{\nu}+\lambda_{\mu}\xi_{\nu}\right)$.

As in the case of the pp-wave spacetimes, the AdS-plane wave also
satisfies special algebraic properties. However, instead of the Riemann
and Ricci tensors, the Weyl tensor and the traceless-Ricci tensor,
that is $S_{\mu\nu}\equiv R_{\mu\nu}-\frac{R}{D}g_{\mu\nu}$, are
Type N. By using the results in \cite{gurses1}, the traceless-Ricci
and Weyl tensors can be calculated as 
\begin{equation}
S_{\mu\nu}=\rho\lambda_{\mu}\lambda_{\nu},\label{eq:Smn_KSK}
\end{equation}
and 
\begin{equation}
C_{\mu\alpha\nu\beta}=4\lambda_{[\mu}\Omega_{\alpha][\beta}\lambda_{\nu]},\label{eq:Weyl_KSK}
\end{equation}
\textit{\emph{where the square brackets denote anti-symmetrization,
and $\rho$ is defined as 
\begin{equation}
\rho\equiv-\left(\square+2\xi^{\mu}\partial_{\mu}+\frac{1}{2}\xi^{\mu}\xi_{\mu}-\frac{2\left(D-2\right)}{\ell^{2}}\right)V,
\end{equation}
and the symmetric tensor $\Omega_{\alpha\beta}$ is defined as 
\begin{equation}
\Omega_{\alpha\beta}\equiv-\left[\nabla_{\alpha}\partial_{\beta}V+\xi_{(\alpha}\partial_{\beta)}V+\frac{1}{2}\xi_{\alpha}\xi_{\beta}V+\frac{1}{D-2}g_{\alpha\beta}\left(\rho-\frac{2\left(D-2\right)}{\ell^{2}}V\right)\right].
\end{equation}
In fact, these forms follow from (\ref{eq:AdS-wave_KS}) and (\ref{eq:AdS-wave_prop}),
and the derivations are given in the App.~\ref{sec:KSK_tensors}.}}
In the given forms above, Type-N properties of the Weyl and traceless-Ricci
tensors are explicit. It can also be seen that the $\lambda^{\mu}$
contractions with the traceless-Ricci tensor is zero. This is also
the case for the Weyl tensor, since $\Omega_{\alpha\beta}$ satisfies
\begin{equation}
\lambda^{\alpha}\Omega_{\alpha\beta}=\frac{1}{2}\lambda_{\beta}\Omega_{\alpha}^{\alpha},
\end{equation}
where 
\begin{equation}
\Omega_{\alpha}^{\alpha}=\xi^{\alpha}\partial_{\alpha}V-\frac{2}{D-2}\rho+\frac{4}{\ell^{2}}V.
\end{equation}
The scalar curvature for the AdS-plane waves is constant $R=-D\left(D-1\right)/\ell^{2}$.
In addition, these spacetimes have constant scalar invariants (CSI),
for example 
\begin{equation}
R_{\mu\alpha\beta\gamma}R^{\mu\alpha\beta\gamma}=\frac{2D\left(D-1\right)}{\ell^{4}},\qquad R_{\mu\sigma}R^{\mu\sigma}=\frac{D\left(D-1\right)^{2}}{\ell^{4}}.
\end{equation}
Lastly, due to $\nabla_{\mu}\lambda_{\nu}=\xi_{(\mu}\lambda_{\nu)}$
and $\lambda^{\mu}\xi_{\mu}=0$, the $\lambda^{\mu}$ vector is nonexpanding,
shear-free, and nontwisting; therefore, the AdS-plane wave metrics
belong to the Kundt class of metrics.

Like pp-wave spacetimes, the two tensors of AdS-plane wave spacetimes
also have a special structure: In \cite{Gurses-PRL}, sketching a
proof using the boost weight decomposition \cite{BoostWeight}, we
gave the following theorem: 
\begin{quotation}
\textit{Consider a Kundt spacetime for which the Weyl and the traceless
Ricci tensors be of type-N, and all scalar invariants be constant.
Then, any second rank symmetric tensor constructed from the Riemann
tensor and its covariant derivatives can be written as a linear combination
of $g_{\mu\nu}$, $S_{\mu\nu}$, and higher orders of $S_{\mu\nu}$(such
as, for example, $\square^{n}\, S_{\mu\nu}$). } 
\end{quotation}
The AdS-plane wave spacetimes belong to this class. Below, we give
a direct proof of this theorem specific to the AdS-plane waves.

\subsection{Two-tensors in AdS-plane wave spacetime}

A generic two tensor \textit{\emph{obtained by contracting any number
of Riemann tensor and its covariant derivatives}} can be symbolically
written as 
\begin{equation}
\left[R^{r_{0}}\left(\nabla^{r_{1}}R\right)\left(\nabla^{r_{2}}R\right)\dots\left(\nabla^{r_{s}}R\right)\right]_{\mu\nu},\label{eq:Generic_two_tensor_of_Riem}
\end{equation}
where the same conventions in the pp-wave case are used. Since the
Riemann tensor is 
\begin{equation}
R_{\mu\alpha\nu\beta}=C_{\mu\alpha\nu\beta}+\frac{2}{D-2}\left(g_{\mu[\nu}S_{\beta]\alpha}-g_{\alpha[\nu}S_{\beta]\mu}\right)+\frac{2R}{D\left(D-1\right)}g_{\mu[\nu}g_{\beta]\alpha},\label{eq:Riemann_in_CSR}
\end{equation}
equivalently, one can write (\ref{eq:Generic_two_tensor_of_Riem})
as a sum of \textit{\emph{terms in the form 
\begin{equation}
\left[C^{m_{0}}\left(\nabla^{m_{1}}C\right)\left(\nabla^{m_{2}}C\right)\dots\left(\nabla^{m_{k}}C\right)S^{n_{0}}\left(\nabla^{n_{1}}S\right)\left(\nabla^{n_{2}}S\right)\dots\left(\nabla^{n_{l}}S\right)\right]_{\mu\nu},\label{eq:Generic_two_tensor}
\end{equation}
where $C$ and $S$ represent the Weyl tensor and the traceless-Ricci
tensor, respectively. }}Note that \textit{\emph{one may consider adding
the metric to (\ref{eq:Generic_two_tensor}) to make the discussion
more complete, but it would be a trivial addition which would boil
down to (\ref{eq:Generic_two_tensor}) after carrying out contractions
involving $g_{\mu\nu}$'s.}}

Here is what we will prove: \emph{The generic two tensors of the form
(\ref{eq:Generic_two_tensor}) will boil down to a linear combination
of $S_{\mu\nu}$ and }\textit{\emph{$\square^{n}\, S_{\mu\nu}$'}}\textit{s}\textit{\emph{.}}

\textit{\emph{The proof is somewhat lengthy and lasts until the end
of this section. The reader who is not interested in the proof, but
just in the applications of the result can skip this section. Now,
let us give the proof which involves two steps:}} 
\begin{enumerate}
\item \textit{\emph{Proving $\left[C^{m_{0}}\left(\nabla^{m_{1}}C\right)\left(\nabla^{m_{2}}C\right)\dots\left(\nabla^{m_{k}}C\right)S^{n_{0}}\left(\nabla^{n_{1}}S\right)\left(\nabla^{n_{2}}S\right)\dots\left(\nabla^{n_{l}}S\right)\right]_{\mu\nu}=0$
unless $\left(m_{0},k,n_{0},l\right)=\left(0,1,0,0\right)$, or $\left(m_{0},k,n_{0},l\right)=\left(0,0,1,0\right)$
or $\left(m_{0},k,n_{0},l\right)=\left(0,0,0,1\right)$.}} 
\item \textit{\emph{For even}}%
\footnote{\textit{\emph{Note that for odd $n$, it is not possible to have a
two tensor contraction.}}%
}\textit{\emph{ $n$, proving $\left[\nabla^{n}S\right]_{\mu\nu}$
and $\left[\nabla^{n}C\right]_{\mu\nu}$ have a second rank tensor
contraction which is a linear combination of $\square^{\frac{n}{2}}S$,
$\square^{\frac{n}{2}-1}S$, $\dots$, $\square S$, and $S$.}} 
\end{enumerate}

\subsubsection{\emph{$\left[C^{m_{0}}\left(\nabla^{m_{1}}C\right)\left(\nabla^{m_{2}}C\right)\dots\left(\nabla^{m_{k}}C\right)S^{n_{0}}\left(\nabla^{n_{1}}S\right)\left(\nabla^{n_{2}}S\right)\dots\left(\nabla^{n_{l}}S\right)\right]_{\mu\nu}=0$
if $m_{0}\protect\ne0$ and $n_{0}+k+l>1$:}}

\textit{\emph{Before giving the precise proof, let us present the
basic idea. If one considers the forms of the Weyl tensor and the
traceless-Ricci tensor together with the property $\nabla_{\mu}\lambda_{\nu}=\xi_{(\mu}\lambda_{\nu)}$,
then one can see that the generic term (\ref{eq:Generic_two_tensor})
represents sum of terms that are made up of $2\left(m_{0}+k+n_{0}+l\right)$
number of $\lambda$ vectors and various combinations of the derivatives
of $V$, the $\xi$ vector and its derivatives. Without loss of generality,
one can assume $m_{1}<m_{2}<\dots<m_{k}$ and $n_{1}<n_{2}<\dots<n_{l}$,
then the building blocks of (\ref{eq:Generic_two_tensor}) are 
\[
\lambda,\,\xi,\,\nabla^{p}V,\,\nabla^{r}\xi,\qquad p=1,\dots,\max\left(n_{l},m_{k}\right)+2;\quad r=1,\dots,\max\left(n_{l},m_{k}\right).
\]
We proved that these building blocks (other than $\lambda$) generate
a free-index $\lambda$ vector when they are contracted with a $\lambda$
vector. In addition, the remaining tensor structure just involves
the same buildings blocks that have the same or lower derivative order
than the order before contraction. Naturally, any tensor that is made
up of these building blocks inherit this property. Due to this property,
it is not possible to lower the number of $\lambda$ vectors by contractions
and these $\lambda$ vectors sooner or later yield a zero contraction.
Therefore, to have a nonzero term out of (\ref{eq:Generic_two_tensor}),
the unique possibility is to have at most two $\lambda$ vectors,
that is 
\[
2\left(m_{0}+k+n_{0}+l\right)=2\Rightarrow m_{0}+k+n_{0}+l=1,
\]
yielding either $\left(m_{0},k,n_{0},l\right)=\left(1,0,0,0\right)$,
$\left(m_{0},k,n_{0},l\right)=\left(0,1,0,0\right)$, $\left(m_{0},k,n_{0},l\right)=\left(0,0,1,0\right)$,
or $\left(m_{0},k,n_{0},l\right)=\left(0,0,0,1\right)$. But, $\left(m_{0},k,n_{0},l\right)=\left(1,0,0,0\right)$
is $\left[C\right]_{\mu\nu}$ which is just zero. Thus, the possible
nonzero terms out of (\ref{eq:Generic_two_tensor}) are in the form
$S_{\mu\nu}$, $\left[\nabla^{n}S\right]_{\mu\nu}$, and $\left[\nabla^{m}C\right]_{\mu\nu}$
which are studied in Sec.~\ref{sub:Reduction-of-nonzero-terms}.}}

Now, let us start our rigorous proof and first show how a free index
$\lambda$ vector is generated by any $\lambda$ contraction. To this
end, consider the behavior of the $\left(0,n\right)$ rank tensor
$\nabla^{n-1}\xi$ under $\lambda$ contractions.\textit{\emph{ To
analyze }}$\nabla^{n-1}\xi$\textit{\emph{, we work in the null frame
in which the metric has the form 
\begin{equation}
ds^{2}=\frac{\ell^{2}}{z^{2}}\left(2dudv+d\vec{x}\cdot d\vec{x}+dz^{2}\right)+2V\left(u,\vec{x},z\right){\rm d}u^{2},\label{eq:AdS-plane_wave_metric}
\end{equation}
}}where $u$ and $v$ are null coordinates, and \textit{\emph{$\vec{x}=\left(x^{i}\right)$
with $i=1,\dots,D-3$}}. Thus, $\lambda_{\mu}$ and $\lambda^{\mu}$
are of the form 
\begin{equation}
\lambda_{\mu}dx^{\mu}=du,\qquad\lambda^{\mu}\partial_{\mu}=\frac{z^{2}}{\ell^{2}}\partial_{v},
\end{equation}
showing why the metric function $V$ does not depend\textit{\emph{
on the coordinate $v$ due to the relation $\lambda^{\mu}\partial_{\mu}V=0$.
In addition, $\xi_{\mu}$ and $\xi^{\mu}$ become \cite{gurses1};
\begin{equation}
\xi_{\mu}=\frac{2}{z}\delta_{\mu}^{z},\qquad\xi^{\mu}=\frac{2z}{\ell^{2}}\delta_{z}^{\mu}.\label{eq:ksi_in_null-frame}
\end{equation}
The property $\xi^{\mu}\lambda_{\mu}=0$ and }}$\left[\nabla_{\mu},\nabla_{\alpha}\right]\lambda^{\mu}=-\frac{\left(D-1\right)}{\ell^{2}}\lambda_{\alpha}$
yield the following identities for the $\xi^{\mu}$ vector; 
\begin{equation}
\lambda^{\mu}\nabla_{\alpha}\xi_{\mu}=-\frac{2}{\ell^{2}}\lambda_{\alpha},\label{eq:ksi_identity_kl0}
\end{equation}
and 
\begin{equation}
\lambda^{\mu}\nabla_{\mu}\xi_{\alpha}=-\frac{2}{\ell^{2}}\lambda_{\alpha},\label{eq:ksi_identity_Ricci-id}
\end{equation}
where we also used $\nabla_{\mu}\xi^{\mu}=-\frac{2\left(D-1\right)}{\ell^{2}}$.

Now, let us look at $\nabla^{n-1}\xi$ in the explicit form; 
\begin{align}
\nabla_{\mu_{1}}\nabla_{\mu_{2}}\dots\nabla_{\mu_{n-1}}\xi_{\mu_{n}}= & \partial_{\mu_{1}}\partial_{\mu_{2}}\dots\partial_{\mu_{n-1}}\xi_{\mu_{n}}-\left(\partial_{\mu_{1}}\partial_{\mu_{2}}\dots\partial_{\mu_{n-2}}\Gamma_{\mu_{n-1}\mu_{n}}^{\sigma_{1}}\right)\xi_{\sigma_{1}}\nonumber \\
 & -\Gamma_{\mu_{n-1}\mu_{n}}^{\sigma_{1}}\partial_{\mu_{1}}\partial_{\mu_{2}}\dots\partial_{\mu_{n-2}}\xi_{\sigma_{1}}\label{eq:Exp_ksi_AdS-wave}\\
 & -\dots-\left(-1\right)^{n-1}\Gamma_{\mu_{1}\mu_{2}}^{\sigma_{1}}\Gamma_{\sigma_{1}\mu_{3}}^{\sigma_{2}}\dots\Gamma_{\sigma_{n-2}\mu_{n}}^{\sigma_{n-1}}\xi_{\sigma_{n-1}}.\nonumber 
\end{align}
The structures appearing in this expression are the Christoffel connection,
partial derivatives of both $\xi_{\mu}$ and the Christoffel connection.
In considering possible $\lambda^{\mu}$ contractions with the terms
in this expansion, first note that (\ref{eq:ksi_in_null-frame}) yields
\begin{equation}
\lambda^{\mu_{n}}\partial_{\mu_{1}}\partial_{\mu_{2}}\dots\partial_{\mu_{n-1}}\xi_{\mu_{n}}=0,\qquad\lambda^{\mu_{j}}\partial_{\mu_{1}}\dots\partial_{\mu_{j}}\dots\partial_{\mu_{n-1}}\xi_{\mu_{n}}=0.\label{eq:lambda_contracts_ksi_partials}
\end{equation}
In addition, since $\partial_{v}g_{\alpha\beta}=0$, a $\lambda^{\mu}$
contraction with the derivatives acting on a Christoffel connection
also yields zero; 
\begin{equation}
\lambda^{\mu_{j}}\partial_{\mu_{1}}\dots\partial_{\mu_{j}}\dots\partial_{\mu_{n-2}}\Gamma_{\mu_{n-1}\mu_{n}}^{\sigma_{1}}=0.\label{eq:lambda_contracts_Chris_partials}
\end{equation}
Moving to the $\lambda^{\mu}$ and $\lambda_{\mu}$ contractions of
the Christoffel connection, the property $\nabla_{\alpha}\lambda_{\beta}=\xi_{(\alpha}\lambda_{\beta)}$
leads to 
\begin{equation}
\Gamma_{\alpha\beta}^{\sigma}\lambda_{\sigma}=-\frac{1}{2}\left(\lambda_{\alpha}\xi_{\beta}+\xi_{\alpha}\lambda_{\beta}\right),\label{eq:lambda_contracts_Chris-I}
\end{equation}
\begin{equation}
\Gamma_{\beta\sigma}^{\alpha}\lambda^{\sigma}=\frac{1}{2}\left(\xi^{\alpha}\lambda_{\beta}-\lambda^{\alpha}\xi_{\beta}\right),\label{eq:lambda_contracts_Chris-II}
\end{equation}
in the null frame. In addition, when $\lambda^{\mu}$ or $\lambda_{\mu}$
contracts with a Christoffel connection under the action of the partial
derivatives, one has 
\begin{equation}
\lambda_{\sigma}\partial_{\mu_{1}}\partial_{\mu_{2}}\dots\partial_{\mu_{n}}\Gamma_{\alpha\beta}^{\sigma}=-\frac{1}{2}\lambda_{\alpha}\partial_{\mu_{1}}\partial_{\mu_{2}}\dots\partial_{\mu_{n}}\xi_{\beta}-\frac{1}{2}\lambda_{\beta}\partial_{\mu_{1}}\partial_{\mu_{2}}\dots\partial_{\mu_{n}}\xi_{\alpha},\label{eq:lambda_contracts_partial_Chris-I}
\end{equation}
\begin{equation}
\lambda^{\sigma}\partial_{\mu_{1}}\partial_{\mu_{2}}\dots\partial_{\mu_{n}}\Gamma_{\beta\sigma}^{\alpha}=\frac{1}{2}\left(\xi^{\alpha}\lambda_{\beta}-\lambda^{\alpha}\xi_{\beta}\right)z\partial_{\mu_{1}}\partial_{\mu_{2}}\dots\partial_{\mu_{n}}\frac{1}{z},\label{eq:lambda_contracts_partial_Chris-II}
\end{equation}
where a new structure, that is partial derivatives acting on $1/z$,
appears; however, it yields zero after a further $\lambda^{\mu}$
contraction.

After discussing all possible $\lambda$ contraction patterns (\ref{eq:lambda_contracts_ksi_partials}--\ref{eq:lambda_contracts_partial_Chris-II})
with the structures involved in the expansion of $\nabla^{n-1}\xi$,
now let us show that a $\lambda$ vector contraction with the $\left(0,n\right)$
rank tensor $\nabla^{n-1}\xi$ provides a free-index $\lambda$ one-form
(we mean $\lambda_{\mu}$). To see this, first notice that the possible
nonzero contractions of the $\lambda$ vector (we mean $\lambda^{\mu}$),
that are (\ref{eq:lambda_contracts_Chris-II}, \ref{eq:lambda_contracts_partial_Chris-II}),
always consist of two terms such that one of them involves a $\lambda$
one-form and the other involves a $\lambda$ vector. If the reproduced
$\lambda$ one-form is noncontracting, then we have achieved the goal
of having a free-index $\lambda$ one-form. However, if it is contracting,
then it must make a contraction in the form of either (\ref{eq:lambda_contracts_Chris-I})
or (\ref{eq:lambda_contracts_partial_Chris-I}), so this contracted
$\lambda$ one-form generates new $\lambda$ one-forms. The same procedure
holds for these newly generated $\lambda$ one-forms and when all
the possible $\lambda$ one-form contractions are carried out, one
always ends up with a free-index $\lambda$ one-form. On the other
hand, returning to the $\lambda$ vector reproduced after the first
contraction, it necessarily makes a contraction and if this contraction
is not zero, it should be again in the form of either (\ref{eq:lambda_contracts_Chris-II})
or (\ref{eq:lambda_contracts_partial_Chris-II}). Thus, one should
follow the same procedure until the newly generated $\lambda$ vector
makes a zero contraction and this is in fact the case since for a
$\lambda$ vector, there is a limited number of nonzero contraction
possibilities generating a new $\lambda$ vector in each term in the
$\nabla^{n-1}\xi$ expansion (\ref{eq:Exp_ksi_AdS-wave}).

For any number of $\lambda^{\mu}$ contractions with the $\left(0,n\right)$
rank tensor $\nabla^{n-1}\xi$, the case is the same and each $\lambda^{\mu}$
contraction generates a free-index $\lambda_{\mu}$ one-form in each
term in the $\nabla^{n-1}\xi$ expansion (\ref{eq:Exp_ksi_AdS-wave})
if it makes a nonzero contraction. To see this, just note that after
each $\lambda^{\mu}$ contraction the remaining structures are the
original ones (the $\xi_{\mu}$ one-form, the Christoffel connection,
partial derivatives of both $\xi_{\mu}$ one-form and the Christoffel
connection) in addition to the newly generated $\xi^{\mu}$ vectors
and $\partial_{\mu_{1}}\partial_{\mu_{2}}\dots\partial_{\mu_{n}}\left(1/z\right)$
type forms which yield zero under a further $\lambda^{\mu}$ contraction.
Therefore, the discussion for the further $\lambda^{\mu}$ contractions
is not different from the single $\lambda^{\mu}$ contraction, and
each $\lambda^{\mu}$ contraction generates a free-index $\lambda$
one-form.

Moving to the other tensor structure appearing in (\ref{eq:Generic_two_tensor}),
the $\left(0,n\right)$ rank tensor $\nabla^{n}V$ also shares the
same properties as $\nabla^{n-1}\xi$ under $\lambda^{\mu}$ contractions.
Expanding $\nabla^{n}V$ yields 
\begin{align}
\nabla_{\mu_{1}}\nabla_{\mu_{2}}\dots\nabla_{\mu_{n-1}}\partial_{\mu_{n}}V= & \partial_{\mu_{1}}\partial_{\mu_{2}}\dots\partial_{\mu_{n-1}}\partial_{\mu_{n}}V-\left(\partial_{\mu_{1}}\partial_{\mu_{2}}\dots\partial_{\mu_{n-2}}\Gamma_{\mu_{n-1}\mu_{n}}^{\sigma_{1}}\right)\partial_{\sigma_{1}}V\nonumber \\
 & -\Gamma_{\mu_{n-1}\mu_{n}}^{\sigma_{1}}\partial_{\mu_{1}}\partial_{\mu_{2}}\dots\partial_{\mu_{n-2}}\partial_{\sigma_{1}}V\label{eq:Exp_V_AdS-wave}\\
 & -\dots-\left(-1\right)^{n-1}\Gamma_{\mu_{1}\mu_{2}}^{\sigma_{1}}\Gamma_{\sigma_{1}\mu_{3}}^{\sigma_{2}}\dots\Gamma_{\sigma_{n-2}\mu_{n}}^{\sigma_{n-1}}\partial_{\sigma_{n-1}}V,\nonumber 
\end{align}
where simply $\partial_{\mu}V$ replaces $\xi_{\mu}$ of the above
discussion. When this expansion is contracted with the $\lambda^{\mu}$
vectors, the possible contraction patterns are the same as the $\nabla^{n-1}\xi$
case except the $\partial_{\mu_{1}}\partial_{\mu_{2}}\dots\partial_{\mu_{n-1}}\xi_{\mu_{n}}$
term is replaced by $\partial_{\mu_{1}}\partial_{\mu_{2}}\dots\partial_{\mu_{n}}V$
which also yields a zero under a $\lambda^{\mu}$ contraction as 
\begin{equation}
\lambda^{\mu_{j}}\partial_{\mu_{1}}\dots\partial_{\mu_{j}}\dots\partial_{\mu_{n}}V=0.
\end{equation}
Therefore, after exactly the same discussion as in the case of $\nabla^{n-1}\xi$,
one can show that each $\lambda^{\mu}$ contraction with $\nabla^{n}V$
generates a free-index $\lambda_{\mu}$ one-form.

We established that each $\lambda$ vector contraction with the $\left(0,n\right)$
rank tensors $\nabla^{n-1}\xi$ and $\nabla^{n}V$ generates a free-index
$\lambda$ one-form; however, after a certain number of $\lambda$
vector contractions, these tensors become necessarily zero. Because
the possible nonzero $\lambda$ vector contractions are made with
the indices of the Christoffel connections and the maximum number
of Christoffel connections is just $\left(n-1\right)$ for both cases.
These $\left(n-1\right)$ number of Christoffel connections involve
$n$ number of \emph{free} down indices. Each $\lambda$ vector contraction
reduces the number of contractable down indices%
\footnote{The ones giving a nonzero result.%
} by $2$ since it also introduces a free-index $\lambda$ one-form.
Thus, if $n$ is even, then $n/2$ is the maximum number of $\lambda$
vector contractions before getting necessarily a zero. On the other
hand, for odd $n$, $\left(n-1\right)/2$ is the maximum number of
nonzero $\lambda$ vector contractions.

In obtaining a two-tensor from the rank $\left(0,4\left(m_{0}+k\right)+2\left(n_{0}+l\right)+\sum_{i=1}^{k}m_{i}+\sum_{i=1}^{l}n_{i}\right)$
tensor 
\begin{equation}
\left[C^{m_{0}}\left(\nabla^{m_{1}}C\right)\left(\nabla^{m_{2}}C\right)\dots\left(\nabla^{m_{k}}C\right)S^{n_{0}}\left(\nabla^{n_{1}}S\right)\left(\nabla^{n_{2}}S\right)\dots\left(\nabla^{n_{l}}S\right)\right],\label{eq:Generic_term}
\end{equation}
one may prefer to make contractions involving $\lambda$ one-forms
first. To have a nonzero contraction, $\lambda$ one-forms should
be contracted with either $\nabla^{n-1}\xi$ tensors or $\nabla^{n}V$
tensors. However, since these contractions generate new $\lambda$
one-forms, the number of $\lambda$ one-forms cannot be reduced by
contractions. In addition, there is a limit for getting a nonzero
contraction out of tensors $\nabla^{n-1}\xi$ and $\nabla^{n}V$.
As a result, in the presence of more than two $\lambda$ one-forms,
one cannot get rid off these $\lambda$ one-forms by contraction and
they, sooner or later, make zero contractions.

To have a nonzero two-tensor out of (\ref{eq:Generic_term}), there
should be at most two $\lambda$ one-forms and they should provide
the two-tensor structure. Then, the possibilities are 
\begin{equation}
S_{\mu\nu},\qquad\left[\nabla^{n}C\right]_{\mu\nu},\qquad\left[\nabla^{n}S\right]_{\mu\nu},
\end{equation}
where we have not included $\left[C\right]_{\mu\nu}$ as the Weyl
tensor is traceless. Note that to have a two-tensor, the number of
covariant derivatives acting on the Weyl tensor and the traceless-Ricci
tensor should be even. Next, we will reduce the last two expressions
into the desired form.

\subsubsection{\emph{Reduction of $\left[\nabla^{n}S\right]_{\mu\nu}$ and $\left[\nabla^{n}C\right]_{\mu\nu}$
to $\sum_{i=0}^{\frac{n}{2}}d_{i}\left(D,R\right)\square^{i}S$:\label{sub:Reduction-of-nonzero-terms}}}

First, let us analyze the \textit{\emph{$\left[\nabla^{n}S\right]_{\mu\nu}$
term where $n$ is even as we want to have a two-tensor contraction
out of $\nabla^{n}S$. The lowest order term is $\left[\nabla^{2}S\right]_{\mu\nu}$
which has two contraction possibilities $\square S_{\mu\nu}$ and
$\nabla^{\alpha}\nabla_{\mu}S_{\alpha\nu}$. The first contraction
possibility is already in the desired form. For the second possibility,
changing the orders of the covariant derivatives yields 
\begin{equation}
\nabla^{\alpha}\nabla_{\mu}S_{\alpha\nu}=\nabla_{\mu}\nabla^{\alpha}S_{\alpha\nu}+\left[\nabla^{\alpha},\nabla_{\mu}\right]S_{\alpha\nu},
\end{equation}
where the first term is zero due to Bianchi identity and the constancy
of the scalar curvature, and finally it takes the form 
\begin{equation}
\nabla^{\alpha}\nabla_{\mu}S_{\alpha\nu}=\frac{R}{D-1}S_{\mu\nu},
\end{equation}
after using (\ref{eq:Riemann_in_CSR}). After discussing the lowest
order, to use mathematical induction, let us analyze a generic $n^{\text{th}}$
order derivative term $\left[\nabla^{n}S\right]_{\mu\nu}$. The contraction
patterns for this term are: first, the two free indices can be on
the $S$ tensor, and secondly, at least one of the free indices is
on the covariant derivatives. In the first contraction pattern, the
indices of the covariant derivatives are totally contracted among
themselves and it is possible to rearrange the order of covariant
derivatives to put the term in the form $\square^{\frac{n}{2}}S_{\mu\nu}$
by using 
\begin{align}
\nabla_{\mu_{1}}\nabla_{\mu_{2}}\left(\prod_{i=3}^{r-2}\nabla_{\mu_{i}}\right)S_{\mu_{r-1}\mu_{r}}= & \nabla_{\mu_{2}}\nabla_{\mu_{1}}\left(\prod_{i=3}^{r-2}\nabla_{\mu_{i}}\right)S_{\mu_{r-1}\mu_{r}}+\left[\nabla_{\mu_{1}},\nabla_{\mu_{2}}\right]\left(\prod_{i=3}^{r-2}\nabla_{\mu_{i}}\right)S_{\mu_{r-1}\mu_{r}}\nonumber \\
= & \nabla_{\mu_{2}}\nabla_{\mu_{1}}\left(\prod_{i=3}^{r-2}\nabla_{\mu_{i}}\right)S_{\mu_{r-1}\mu_{r}}\nonumber \\
 & +\sum_{s=3}^{r-2}R_{\mu_{1}\mu_{2}\mu_{s}}^{\phantom{\mu_{1}\mu_{2}\mu_{s}}\mu_{r+1}}\left(\prod_{i_{1}=3}^{s-1}\nabla_{\mu_{i_{_{1}}}}\right)\nabla_{\mu_{r+1}}\left(\prod_{i_{2}=s+1}^{r-2}\nabla_{\mu_{i_{_{2}}}}\right)S_{\mu_{r-1}\mu_{r}}\nonumber \\
 & +R_{\mu_{1}\mu_{2}\mu_{r-1}}^{\phantom{\mu_{1}\mu_{2}\mu_{r-1}}\mu_{r+1}}\left(\prod_{i=3}^{r-2}\nabla_{\mu_{i}}\right)S_{\mu_{r+1}\mu_{r}}\nonumber \\
 & +R_{\mu_{1}\mu_{2}\mu_{r}}^{\phantom{\mu_{1}\mu_{2}\mu_{r}}\mu_{r+1}}\left(\prod_{i=3}^{r-2}\nabla_{\mu_{i}}\right)S_{\mu_{r-1}\mu_{r+1}}.\label{eq:Order_change}
\end{align}
In addition, if one uses (\ref{eq:Riemann_in_CSR}), then the parts
of the Riemann tensor involving the Weyl and the traceless-Ricci tensors
just yield zeros as we proved above. The remaining nonzero part of
the Riemann tensor in which the tensor structure is just two metrics
{[}that is the third term in (\ref{eq:Riemann_in_CSR}){]} reduces
the terms involving the Riemann tensor to $\left(n-2\right)^{{\rm th}}$
order terms as $\left[\nabla^{n-2}S\right]_{\mu\nu}$. Thus, first
contraction pattern of $\left[\nabla^{n}S\right]_{\mu\nu}$ yields
a sum involving $\square^{\frac{n}{2}}S_{\mu\nu}$ and $\left[\nabla^{n-2}S\right]_{\mu\nu}$
terms. On the other hand, for the second contraction pattern, at least
one of the covariant derivatives is contracted with $S$ and to use
the Bianchi identity $\nabla^{\rho}S_{\mu\rho}=0$, one needs to change
the order of the covariant derivative contracting with $S$ until
placing it next to $S$ by using (\ref{eq:Order_change}). During
this process again terms involving the Riemann tensor and $\left(n-2\right)$
number of covariant derivatives are introduced, and after use of (\ref{eq:Riemann_in_CSR}),
these terms become $\left[\nabla^{n-2}S\right]_{\mu\nu}$. Thus, the
second contraction pattern of $\left[\nabla^{n}S\right]_{\mu\nu}$
yields a sum involving $\left[\nabla^{n-2}S\right]_{\mu\nu}$ terms.
Then, as we showed that the lowest order derivative term $\left[\nabla^{2}S\right]_{\mu\nu}$
satisfies the desired pattern and that the $n^{{\rm th}}$ order term
$\left[\nabla^{n}S\right]_{\mu\nu}$ reduces to a sum involving a
desired term $\square^{\frac{n}{2}}S_{\mu\nu}$ and $\left(n-2\right)^{{\rm th}}$
order terms $\left[\nabla^{n-2}S\right]_{\mu\nu}$'s, it is clear
by mathematical induction that $\left[\nabla^{n}S\right]_{\mu\nu}$
can be represented as a sum in the form 
\begin{equation}
\left[\nabla^{n}S\right]_{\mu\nu}=\sum_{i=0}^{\frac{n}{2}}d_{i}\left(D,R\right)\square^{i}S_{\mu\nu},\label{eq:Nonzero_S_derivative_terms}
\end{equation}
where $d_{n/2}$ is just one, and the dimension and scalar curvature
dependence of other $d_{i}$'s are due to the Riemann tensors that
are transformed via (\ref{eq:Riemann_in_CSR}).}}

Now, let us move to the term \textit{\emph{$\left[\nabla^{n}C\right]_{\mu\nu}$
where $n$ is again even as we want to have a two-tensor contraction
out of $\nabla^{n}C$. Since the Weyl tensor is traceless, at least
two covariant derivatives should be contracted with the Weyl tensor
in obtaining a nonzero two tensor from $\nabla^{n}C$. Then, at the
lowest order $\left[\nabla^{2}C\right]_{\mu\nu}=\nabla^{\alpha}\nabla^{\beta}C_{\mu\alpha\nu\beta}$,
one can use the following identity for the Weyl tensor assuming that
the metric is (\ref{eq:AdS-wave_KS}) 
\begin{equation}
\nabla^{\mu}\nabla^{\nu}C_{\mu\alpha\nu\beta}=\frac{D-3}{D-2}\left(\square S_{\alpha\beta}-\frac{R}{D-1}S_{\alpha\beta}\right),\label{eq:Bianchi_Weyl}
\end{equation}
which is derived in the App.~\ref{sec:Bianchi_Weyl}, and then we
immediately obtain the desired form. Now, moving to the $n^{\text{th}}$
order term $\left[\nabla^{n}C\right]_{\mu\nu}$ for which again one
can change the order of covariant derivatives in such a way that two
of the covariant derivatives}}\textit{\textcolor{red}{\emph{ }}}\textit{\emph{contracting
with the Weyl tensor are moved next to it to use the Bianchi identity
(\ref{eq:Bianchi_Weyl}). During the order change of the covariant
derivatives again Riemann tensors are introduced. After use of (\ref{eq:Riemann_in_CSR}),
only the part of the Riemann tensor involving two metrics yields a
nonzero contribution, so the terms involving the Riemann tensor reduce
to $\left(n-2\right)^{{\rm th}}$ order terms $\left[\nabla^{n-2}C\right]_{\mu\nu}$.
Thus, the $n^{\text{th}}$ order term $\left[\nabla^{n}C\right]_{\mu\nu}$
reduces to the sum of $\left[\nabla^{n-2}\square S\right]_{\mu\nu}$,
$\left[\nabla^{n-2}S\right]_{\mu\nu}$, and $\left[\nabla^{n-2}C\right]_{\mu\nu}$
terms. Then, as we showed that the lowest order derivative term $\left[\nabla^{2}C\right]_{\mu\nu}$
can be converted to the $\left[\nabla^{2}S\right]_{\mu\nu}$ case
and that the $n^{{\rm th}}$ order term $\left[\nabla^{n}C\right]_{\mu\nu}$
reduces to a sum involving the $\left[\nabla^{n}S\right]_{\mu\nu}$
and $\left[\nabla^{n-2}S\right]_{\mu\nu}$ cases, and $\left(n-2\right)^{{\rm th}}$
order terms $\left[\nabla^{n-2}C\right]_{\mu\nu}$, it is clear by
mathematical induction that $\left[\nabla^{n}C\right]_{\mu\nu}$ can
be represented as a sum involving just $\left[\nabla^{m}S\right]_{\mu\nu}$
terms where $n\ge m\ge0$. Then, $\left[\nabla^{n}C\right]_{\mu\nu}$
case reduces to the $\left[\nabla^{n}S\right]_{\mu\nu}$ case which
is of the desired form (\ref{eq:Nonzero_S_derivative_terms}).}}

As a result, the nonzero two tensors of the AdS-plane wave spacetime
can be written as a linear combination of the tensor $S_{\mu\nu}$,
$\square^{n}S_{\mu\nu}$'s, and also naturally the metric $g_{\mu\nu}$.
This completes the proof.

Note that with this result about the two-tensors, the CSI property
of the AdS-plane wave spacetimes is explicit since $S_{\mu\nu}$ is
traceless.

\subsection{Field equations of the generic gravity theory for AdS-plane wave
spacetime}

In \cite{Gurses-PRL}, we studied the field equations of the generic
gravity theory for the \textit{\emph{CSI Kundt spacetime of Type-N
Weyl and Type-N traceless Ricci. In addition, we also demonstrated
how the field equations further reduce for }}Kerr-Schild--Kundt spacetimes
to which AdS-plane wave belongs. Let us recapitulate these results
here. \textit{\emph{As an immediate result of our}} conclusions above,
the field equations coming from (\ref{eq:Generic_gravity}) are\textit{\emph{
\begin{equation}
eg_{\mu\nu}+\sum_{n=0}^{N}a_{n}\square^{n}\, S_{\mu\nu}=0.
\end{equation}
The trace of the field equation yields 
\begin{equation}
e=0,
\end{equation}
which determines the effective cosmological constant $\Lambda$ or
$1/\ell^{2}$ in terms of the parameters that appear in the Lagrangian.
On the other hand, the traceless part of the field equation 
\begin{equation}
\sum_{n=0}^{N}a_{n}\square^{n}\, S_{\mu\nu}=0,\label{eq:EoM_gen_grav_as_sum-AdS-plane_wave}
\end{equation}
can be factorized as}} 
\begin{equation}
\prod_{n=1}^{N}\left(\square+b_{n}\right)S_{\mu\nu}=0,\label{eq:Multiplicative_EoM_Smn}
\end{equation}
where $b_{n}$'s are again functions of the parameters of the original
theory, and in general they can be complex which appear in complex
conjugate pairs. To further reduce (\ref{eq:Multiplicative_EoM_Smn}),
note that in \cite{gurses1}, it was shown that for any $\phi$ satisfying
$\lambda^{\mu}\partial_{\mu}\phi=\partial_{v}\phi=0$, one has 
\begin{equation}
\square\phi=\bar{\square}\phi,
\end{equation}
where $\bar{\square}\equiv\bar{g}^{\mu\nu}\bar{\nabla}_{\mu}\bar{\nabla}_{\nu}$.
Therefore, $S_{\mu\nu}=\lambda_{\mu}\lambda_{\nu}\mathcal{O}V$ with
\begin{equation}
\mathcal{O}\equiv-\left(\bar{\square}+2\xi^{\mu}\partial_{\mu}+\frac{1}{2}\xi^{\mu}\xi_{\mu}-\frac{2\left(D-2\right)}{\ell^{2}}\right)=-\left(\bar{\square}+\frac{4z}{\ell^{2}}\partial_{z}-\frac{2\left(D-3\right)}{\ell^{2}}\right),
\end{equation}
where the second equality is valid for AdS-plane wave in the coordinates
(\ref{eq:AdS-plane_wave_metric}). Using the results in \cite{gurses1},
$\square\left(\phi\lambda_{\alpha}\lambda_{\beta}\right)$ can be
written as 
\begin{equation}
\square\left(\phi\lambda_{\alpha}\lambda_{\beta}\right)=\bar{\square}\left(\phi\lambda_{\alpha}\lambda_{\beta}\right)=-\lambda_{\alpha}\lambda_{\beta}\left(\mathcal{O}+\frac{2}{\ell^{2}}\right)\phi,\label{eq:phi_id}
\end{equation}
which is again valid for any $\phi$ satisfying $\lambda^{\mu}\partial_{\mu}\phi=\partial_{v}\phi=0$;
therefore, $\square S_{\mu\nu}$ becomes 
\begin{equation}
\square S_{\mu\nu}=-\lambda_{\mu}\lambda_{\nu}\left(\mathcal{O}+\frac{2}{\ell^{2}}\right)\rho=-\lambda_{\mu}\lambda_{\nu}\left(\mathcal{O}+\frac{2}{\ell^{2}}\right)\mathcal{O}V.\label{eq:Box_Smn}
\end{equation}
Then, (\ref{eq:Multiplicative_EoM_Smn}) becomes 
\begin{equation}
\lambda_{\mu}\lambda_{\nu}\mathcal{O}\prod_{n=1}^{N}\left(\mathcal{O}+\frac{2}{\ell^{2}}-b_{n}\right)V=0,\label{eq:Final_EoM_AdS-wave}
\end{equation}
where we also used the fact that for any $\phi$ satisfying $\partial_{v}\phi=0$,
$\mathcal{O}\phi$ also satisfies the same property $\partial_{v}\mathcal{O}\phi=0$.

Note that (\ref{eq:Final_EoM_AdS-wave}) is linear in the metric function
$V$ which suggests the linearization of the field equations of the
generic theory for the AdS-plane waves. To make this more explicit,
using (\ref{eq:phi_id}) $S_{\mu\nu}$ can be put in the form 
\begin{equation}
S_{\mu\nu}=-\left(\bar{\square}+\frac{2}{\ell^{2}}\right)\left(\lambda_{\mu}\lambda_{\nu}V\right)=-\frac{1}{2}\left(\bar{\square}+\frac{2}{\ell^{2}}\right)h_{\mu\nu},\label{eq:Smn_in_hmn}
\end{equation}
after defining $h_{\mu\nu}\equiv2V\lambda_{\mu}\lambda_{\nu}$ with
which the AdS-plane wave metric becomes $g_{\mu\nu}=\bar{g}_{\mu\nu}+h_{\mu\nu}$.
In addition, using $\partial_{v}\phi=0\Rightarrow\partial_{v}\mathcal{O}\phi=0$
and (\ref{eq:phi_id}), $\square^{n}S_{\mu\nu}$ becomes 
\begin{equation}
\square^{n}S_{\mu\nu}=\left(-1\right)^{n}\lambda_{\mu}\lambda_{\nu}\left(\mathcal{O}+\frac{2}{\ell^{2}}\right)^{n}\mathcal{O}V=\bar{\square}^{n}S_{\mu\nu}.\label{eq:BoxSmn}
\end{equation}
Once (\ref{eq:Smn_in_hmn}) and (\ref{eq:BoxSmn}) are considered
in either (\ref{eq:EoM_gen_grav_as_sum-AdS-plane_wave}) or (\ref{eq:Multiplicative_EoM_Smn}),
it is obvious that the field equations of the generic theory (\ref{eq:Generic_gravity})
for AdS-plane waves are linear in $h_{\mu\nu}$ as in the case of
a perturbative expansion of the field equations around an (A)dS background
for a small metric perturbation $\left\Vert h\right\Vert \equiv\left\Vert g-\bar{g}\right\Vert \ll1$.

As in the case of pp-waves, this observation suggests that there are
two possible ways to find the field equations of the generic gravity
theory for AdS-plane waves: 1- Deriving the field equations and directly
plugging the AdS-plane wave metric ansatz (\ref{eq:AdS-wave_KS})
in them; or 2- Linearizing the derived field equations around the
(A)dS background and putting $h_{\mu\nu}=2V\lambda_{\mu}\lambda_{\nu}$
in these linearized equations. Again, as we discuss in Sec.~\ref{sec:fRiem_solns},
the idea in the second way of finding the field equations for AdS-plane
waves provides a shortcut in finding the field equations of a gravity
theory described with a Lagrangian density which is constructed by
the Riemann tensor but not its derivatives. Lastly, $h_{\mu\nu}=2V\lambda_{\mu}\lambda_{\nu}$
is transverse, $\bar{\nabla}_{\mu}h^{\mu\nu}=0$, and traceless, $\bar{g}^{\mu\nu}h_{\mu\nu}=0$,
so one needs only the linearized field equations for the transverse-traceless
metric perturbation.

Just like the discussion in the pp-wave case, assuming nonvanishing
and distinct $b_{n}$'s, the most general solution of (\ref{eq:Final_EoM_AdS-wave})
is 
\begin{equation}
V=V_{E}+\Re\left(\sum_{n=1}^{N}V_{n}\right),\label{funct}
\end{equation}
where $\Re$ represents the real part and $V_{E}$ is the solution
to the cosmological Einstein's theory, namely 
\begin{equation}
\mathcal{O}V_{E}=-\left(\bar{\square}+\frac{4z}{\ell^{2}}\partial_{z}-\frac{2\left(D-3\right)}{\ell^{2}}\right)V_{E}=-\left(\frac{z^{2}}{\ell^{2}}\hat{\partial}^{2}+\frac{\left(6-D\right)z}{\ell^{2}}\partial_{z}-\frac{2\left(D-3\right)}{\ell^{2}}\right)V_{E}=0,\label{eq:V_E_operators}
\end{equation}
where $\hat{\partial}^{2}\equiv\frac{\partial^{2}}{\partial z^{2}}+\sum_{i=1}^{D-3}\frac{\partial^{2}}{\partial x^{i}\partial x^{i}}$.
Here, the second equality follows from the results in \cite{Gullu-Gurses}.
In addition, $V_{n}$'s solve the equation of the quadratic gravity
theory, i.e. $\left(\mathcal{O}+\frac{2}{\ell^{2}}-b_{n}\right)V_{n}=0$.
As a result, the AdS-plane wave solutions of Einstein's gravity and
quadratic gravity, which were summarized in the Introduction (with
$M_{n}^{2}=-b_{n}+\frac{2}{\ell^{2}}$)%
\footnote{Note that $M_{n}^{2}$ represents the mass of a massive spin-2 excitation
around the AdS background. To prevent any confusion in its definition
observe that $\mathcal{O}\sim-\bar{\square}$. In addition, remember
that in the pp-wave case, we also defined $M_{n,\text{flat}}^{2}$,
the mass around the flat background, and these two masses are related
as $\underset{\ell\rightarrow\infty}{\lim}M_{n}^{2}=M_{n,\text{flat}}^{2}$.%
}, also solves a generic gravity theory \cite{Gurses-PRL}.

When we let $b_{n}=-M_{n}^{2}+\frac{2}{\ell^{2}}$ for all $n=1,2,\cdots,N$
and assume real $M_{n}^{2}$'s as they represent the masses of the
excitations, then we can express the exact solution of the generic
gravity theory depending on $u$ and $z$ as a sum of the Einsteinian
Kaigorodov solution 
\begin{equation}
V_{E}\left(u,z\right)=c\left(u\right)\, z^{D-3},
\end{equation}
where we omitted the $1/z^{2}$ solution as it can be absorbed into
the background AdS metric by the redefinition of the coordinate $v$,
and the functions $V_{n}$'s defined in (\ref{funct}) are given as
\begin{equation}
V_{n}\left(u,z\right)=z^{\frac{D-5}{2}}\left(c_{n,1}\left(u\right)z^{\left|\nu_{n}\right|}+c_{n,2}\left(u\right)z^{-\left|\nu_{n}\right|}\right),~~~~n=1,2,\cdots,N,
\end{equation}
where $\nu_{n}=\frac{1}{2}\sqrt{\left(D-1\right)^{2}+4\ell^{2}M_{n}^{2}}$
and all $M_{n}^{2}$'s are assumed to be distinct. On the other hand,
there can be many special cases in which some of $M_{n}^{2}$'s are
equal. In fact, these special cases do appear in the critical gravity
theories \cite{LuPope,DeserLiu,berg1,Nutma,Pang-BlackHole} and the
corresponding solutions always involve logarithms; for example, for
four-dimensional case see \cite{Alishah,Gullu-Gurses}. Here, let
us mention the extreme case in which all $N$ masses vanish. In this
case, the field equation takes the form 
\begin{equation}
\mathcal{O}^{N+1}V\left(u,z\right)=\left[z^{2}\partial_{z}^{2}+\left(6-D\right)z\partial_{z}-2\left(D-3\right)\right]^{N+1}V\left(u,z\right)=0,
\end{equation}
which has the solution 
\begin{equation}
V\left(u,z\right)=z^{D-3}\,\sum_{n=0}^{N}\, c_{n,1}\left(u\right)\,\ln^{n}z+\frac{1}{z^{2}}\,\sum_{n=1}^{N}\, c_{n,2}\left(u\right)\,\ln^{n}z.
\end{equation}
where again we considered the $1/z^{2}$ solution as absorbed into
AdS part.

\section{pp-Wave and AdS-Plane Wave in Quadratic Gravity\label{sec:Waves_of_quadratic_gravity}}

Since quadratic gravity played a central role in constructing the
solutions of the generic gravity theory, let us explicitly study the
field equations of quadratic gravity in the context of these wave
solutions. The field equations of quadratic gravity \cite{DeserTekin}
\begin{align}
\frac{1}{\kappa}\left(R_{\mu\nu}-\frac{1}{2}g_{\mu\nu}R+\Lambda_{0}g_{\mu\nu}\right)\nonumber \\
+2\alpha R\left(R_{\mu\nu}-\frac{1}{4}g_{\mu\nu}R\right)+\left(2\alpha+\beta\right)\left(g_{\mu\nu}\square-\nabla_{\mu}\nabla_{\nu}\right)R\nonumber \\
+\beta\square\left(R_{\mu\nu}-\frac{1}{2}g_{\mu\nu}R\right)+2\beta\left(R_{\mu\sigma\nu\rho}-\frac{1}{4}g_{\mu\nu}R_{\sigma\rho}\right)R^{\sigma\rho}\nonumber \\
+2\gamma\biggl[RR_{\mu\nu}-2R_{\mu\sigma\nu\rho}R^{\sigma\rho}+R_{\mu\sigma\rho\tau}R_{\nu}^{\phantom{\nu}\sigma\rho\tau}-2R_{\mu\sigma}R_{\nu}^{\phantom{\nu}\sigma}\nonumber \\
-\frac{1}{4}g_{\mu\nu}\left(R_{\tau\lambda\sigma\rho}^{2}-4R_{\sigma\rho}^{2}+R^{2}\right)\biggr] & =0,\label{fieldequations}
\end{align}
for AdS-plane wave (\ref{eq:AdS-wave_KS}) reduce to a trace part
and an apparently nonlinear wave type equation on the traceless-Ricci
tensor \cite{MalekPravda} 
\begin{equation}
\left(\frac{\Lambda_{0}}{\kappa}+\frac{\left(D-1\right)\left(D-2\right)}{2\kappa\ell^{2}}-f\frac{\left(D-1\right)^{2}\left(D-2\right)^{2}}{2\ell^{4}}\right)g_{\mu\nu}+\beta\left(\square+\frac{2}{\ell^{2}}-M^{2}\right)S_{\mu\nu}=0,\label{eq:EoM_quad_AdS-plane}
\end{equation}
where $S_{\mu\nu}$ and $M^{2}$ are given in (\ref{eq:Smn_KSK})
and (\ref{eq:M2}), respectively, and $f$ is 
\begin{equation}
f\equiv\left(D\alpha+\beta\right)\frac{\left(D-4\right)}{\left(D-2\right)^{2}}+\gamma\frac{\left(D-3\right)\left(D-4\right)}{\left(D-1\right)\left(D-2\right)}.
\end{equation}
The trace part of (\ref{eq:EoM_quad_AdS-plane}) 
\begin{equation}
\frac{\Lambda_{0}}{\kappa}+\frac{\left(D-1\right)\left(D-2\right)}{2\kappa\ell^{2}}-f\frac{\left(D-1\right)^{2}\left(D-2\right)^{2}}{2\ell^{4}}=0,\label{eq:AdS_EoM_quad}
\end{equation}
determines the effective cosmological constant, that is the AdS radius
$\ell$. Since $\square S_{\mu\nu}=\bar{\square}S_{\mu\nu}$, the
traceless part of (\ref{eq:EoM_quad_AdS-plane}), upon use of (\ref{eq:phi_id}),
further reduces to 
\begin{equation}
\left(\bar{\square}+\frac{2}{\ell^{2}}-M^{2}\right)\left(\bar{\square}+\frac{2}{\ell^{2}}\right)\left(\lambda_{\mu}\lambda_{\nu}V\right)=0.\label{eq:Linearized_quad}
\end{equation}
This is an exact equation for the AdS-plane waves, but it is also
important to realize that defining $h_{\mu\nu}\equiv g_{\mu\nu}-\bar{g}_{\mu\nu}=2V\lambda_{\mu}\lambda_{\nu}$,
this is also the linearized field equations for transverse-traceless
fluctuations, which represent the helicity $\pm2$ excitations, about
the AdS background whose radius is determined by the trace part of
(\ref{eq:EoM_quad_AdS-plane}).

In this work, we have been interested in the exact solutions and not
perturbative excitations, but as a side remark we can note that the
fact that AdS-plane waves and pp-waves lead to the linearized equations
can be used to put constraints on the original parameters of the theory
once unitarity of the linearized excitations is imposed. For example,
since the excitations cannot be tachyonic or ghost like, $M^{2}\ge0$
and this immediately says that $b_{n}$ cannot be complex.

To obtain the field equations for pp-waves in this theory with $\Lambda_{0}=0$,
one simply takes $\ell\rightarrow\infty$ limit. Note that in this
limit $S_{\mu\nu}$ becomes equal to $R_{\mu\nu}$.

\section{Wave Solutions of $f\left(\text{Riemann}\right)$ Theory\label{sec:fRiem_solns}}

Let us now consider a subclass of the generic theory (\ref{eq:Functional_general_theory})
whose action is built only on the contractions of the Riemann tensor
but not its derivatives. Namely, the action is given as 
\begin{equation}
I=\int d^{D}x\,\sqrt{-g}f\left(R_{\alpha\beta}^{\mu\nu}\right),\label{eq:fRiem_act}
\end{equation}
where we specifically choose $R_{\alpha\beta}^{\mu\nu}\equiv R_{\phantom{\mu\nu}\alpha\beta}^{\mu\nu}$
as the argument to remove the functional dependence on the inverse
metric $g^{\mu\nu}$ without losing any generality. Because any higher
curvature combination can be written in terms of $R_{\alpha\beta}^{\mu\nu}$
without use of either metric or its inverse.

This class of theories constitute an important subclass because of
two facts. First, as we discussed above, pp-waves and AdS-plane waves
(actually, AdS waves in general) linearize the field equations of
a generic gravity theory, that is both plugging the pp-wave (AdS-plane
wave) metric $g_{\mu\nu}=\eta_{\mu\nu}+2V\lambda_{\mu}\lambda_{\nu}$
($g_{\mu\nu}=\bar{g}_{\mu\nu}+2V\lambda_{\mu}\lambda_{\nu}$) in the
field equations and plugging $h_{\mu\nu}=2V\lambda_{\mu}\lambda_{\nu}$
in the linearized field equations around flat (AdS) background yield
the same field equations. Second fact is that for the $f\left(R_{\alpha\beta}^{\mu\nu}\right)$
theory, one can construct a quadratic curvature gravity theory which
has the same vacua and the same linearized field equations as the
original $f\left(R_{\alpha\beta}^{\mu\nu}\right)$ theory (see \cite{Hindawi,BI-Unitarity,AllUni3D,AllUniD,SpectraLovelock,Senturk,UnitaryBI}).
Once one constructs the equivalent quadratic curvature action (EQCA)
corresponding to (\ref{eq:fRiem_act}), by using the effective parameters
of EQCA in the results obtained for the quadratic gravity case in
Sec.~\ref{sec:Waves_of_quadratic_gravity}, one can obtain the field
equations of (\ref{eq:fRiem_act}) for AdS-plane waves and pp-waves
without deriving the field equations of (\ref{eq:fRiem_act}). Use
of EQCA procedure in finding the field equations for AdS-plane waves
and pp-waves provides a fair amount of simplification over the standard
way of finding the field equations which can be quite complicated
depending on function $f$. With a known $f$, to find the corresponding
EQCA, one can use the procedure given in \cite{Senturk} as 
\begin{enumerate}
\item Calculate $f\left(\bar{R}_{\rho\sigma}^{\mu\nu}\right)$, that is
the value of Lagrangian density for the maximally symmetric background
\begin{equation}
\bar{R}_{\rho\sigma}^{\mu\nu}=-\frac{1}{\ell^{2}}\left(\delta_{\rho}^{\mu}\delta_{\sigma}^{\nu}-\delta_{\sigma}^{\mu}\delta_{\rho}^{\nu}\right).\label{eq:Maximally_symmetric_background}
\end{equation}
In addition, take the first and second order derivatives of $f\left(R_{\rho\sigma}^{\mu\nu}\right)$
with respect to the Riemann tensor, and calculate them again for the
background (\ref{eq:Maximally_symmetric_background}) to find 
\begin{align}
\left[\frac{\partial f}{\partial R_{\rho\sigma}^{\mu\nu}}\right]_{\bar{R}_{\rho\sigma}^{\mu\nu}}R_{\rho\sigma}^{\mu\nu} & \equiv\zeta R,\label{eq:First_order}\\
\frac{1}{2}\left[\frac{\partial^{2}f}{\partial R_{\rho\sigma}^{\mu\nu}\partial R_{\lambda\gamma}^{\alpha\beta}}\right]_{\bar{R}_{\rho\sigma}^{\mu\nu}}R_{\rho\sigma}^{\mu\nu}R_{\lambda\gamma}^{\alpha\beta} & \equiv\tilde{\alpha}R^{2}+\tilde{\beta}R_{\nu}^{\mu}R_{\mu}^{\nu}+\tilde{\gamma}\left(R_{\rho\sigma}^{\mu\nu}R_{\mu\nu}^{\rho\sigma}-4R_{\nu}^{\mu}R_{\mu}^{\nu}+R^{2}\right),\label{eq:Second_order}
\end{align}
where $\zeta$, $\tilde{\alpha}$, $\tilde{\beta}$, $\tilde{\gamma}$
are to be determined from these equations. 
\item Construct the action 
\begin{align}
I_{\text{EQCA}}=\int d^{D}x\,\sqrt{-g} & \Biggl[\frac{1}{\tilde{\kappa}}\left(R-2\tilde{\Lambda}_{0}\right)\label{eq:EQCA}\\
 & +\tilde{\alpha}R^{2}+\tilde{\beta}R_{\nu}^{\mu}R_{\mu}^{\nu}+\tilde{\gamma}\left(R_{\rho\sigma}^{\mu\nu}R_{\mu\nu}^{\rho\sigma}-4R_{\nu}^{\mu}R_{\mu}^{\nu}+R^{2}\right)\Biggr].\nonumber 
\end{align}
where $\tilde{\alpha}$, $\tilde{\beta}$, and $\tilde{\gamma}$ will
appear exactly as defined in step 1, while the other remaining two
parameters are given as 
\begin{align}
\frac{1}{\tilde{\kappa}} & =\zeta+\frac{2}{\ell^{2}}\left[\left(D-1\right)\left(D\tilde{\alpha}+\tilde{\beta}\right)+\left(D-2\right)\left(D-3\right)\tilde{\gamma}\right],\label{eq:kappa_eff}\\
\frac{\tilde{\Lambda}_{0}}{\tilde{\kappa}} & =-\frac{1}{2}f\left(\bar{R}_{\rho\sigma}^{\mu\nu}\right)-\frac{D\left(D-1\right)}{2\ell^{2}}\left(\zeta+\frac{1}{\ell^{2}}\left[\left(D-1\right)\left(D\tilde{\alpha}+\tilde{\beta}\right)+\left(D-2\right)\left(D-3\right)\tilde{\gamma}\right]\right).\label{eq:Lambda_0_eff}
\end{align}

\end{enumerate}
After constructing the EQCA corresponding to (\ref{eq:fRiem_act}),
the field equations of (\ref{eq:fRiem_act}) for AdS-plane waves can
be found by replacing the effective parameters of (\ref{eq:EQCA})
in (\ref{eq:EoM_quad_AdS-plane}). Since these field equations are
solved by the AdS-plane wave solutions of Einstein's gravity and quadratic
gravity listed in the Introduction, then the AdS-plane wave solutions
of (\ref{eq:fRiem_act}) simply follow from these solutions by plugging
the effective cosmological constant of (\ref{eq:fRiem_act}) and $M^{2}$
of (\ref{eq:fRiem_act}), which is calculated by putting the effective
parameters of EQCA in (\ref{eq:M2}). The effective cosmological constant
of (\ref{eq:fRiem_act}) can be found from (\ref{eq:AdS_EoM_quad})
after putting the effective parameters of EQCA in. Note that although
(\ref{eq:AdS_EoM_quad}) is, apparently, a quadratic equation in $1/\ell^{2}$,
after putting the effective parameters in this equation it yields
a different dependence on $1/\ell^{2}$ since these effective parameters
also depend on $\ell^{2}$.

As in the case of the quadratic curvature gravity, $\ell\rightarrow\infty$
limit in the AdS-plane wave field equations gives the field equations
for the pp-waves for the theory with $\Lambda_{0}=0$. Equivalently,
one may find the curvature expansion of $f\left(R_{\alpha\beta}^{\mu\nu}\right)$
up to the quadratic order, and this part of the action determines
the field equations for the pp-wave metric.

As an application with a given $f\left(R_{\alpha\beta}^{\mu\nu}\right)$,
we consider the cubic curvature gravity generated by the bosonic string
theory at the second order in inverse string tension $\alpha^{\prime}$
\cite{Tseytlin} and the Lanczos-Lovelock theory \cite{Lanczos,Lovelock}.

\subsection{Cubic gravity generated by string theory}

The effective action for the bosonic string at $O\left[\left(\alpha^{\prime}\right)^{2}\right]$
is 
\begin{align}
I=\frac{1}{\kappa}\int d^{D}x\sqrt{-g} & \Biggl[R+\frac{\alpha^{\prime}}{4}\left(R_{\alpha\beta}^{\mu\nu}R_{\mu\nu}^{\alpha\beta}-4R_{\nu}^{\mu}R_{\mu}^{\nu}+R^{2}\right)\nonumber \\
 & +\frac{\left(\alpha^{\prime}\right)^{2}}{24}\left(-2R_{\nu\beta}^{\mu\alpha}R_{\mu\lambda}^{\nu\gamma}R_{\alpha\gamma}^{\beta\lambda}+R_{\alpha\beta}^{\mu\nu}R_{\mu\nu}^{\gamma\lambda}R_{\gamma\lambda}^{\alpha\beta}\right)\Biggr],\label{eq:Effective_string}
\end{align}
where the bare cosmological constant is not introduced, so the theory
admits a flat background in addition to the (A)dS ones. In \textcolor{blue}{\cite{AllUniD}},
the EQCA of (\ref{eq:Effective_string}) was calculated as 
\begin{align}
I_{\text{EQCA}}=\frac{1}{\kappa}\int d^{D}x\sqrt{-g}\, & \Biggl[\left(1+\frac{\alpha^{\prime2}\left(D-5\right)}{4\ell^{4}}\right)\left(R+2\frac{\alpha^{\prime2}D\left(D-1\right)\left(D-5\right)}{6\left(D-5\right)\alpha^{\prime2}\ell^{2}+24\ell^{6}}\right)\nonumber \\
 & +\frac{\alpha^{\prime2}}{2\ell^{2}}R^{2}-\frac{7\alpha^{\prime2}}{4\ell^{2}}R_{\nu}^{\mu}R_{\mu}^{\nu}+\frac{\alpha^{\prime}}{4}\left(1-\frac{2\alpha^{\prime}}{\ell^{2}}\right)\left(R_{\alpha\beta}^{\mu\nu}R_{\mu\nu}^{\alpha\beta}-4R_{\nu}^{\mu}R_{\mu}^{\nu}+R^{2}\right)\Biggr],\label{eq:eqa_eff_string}
\end{align}
where the effective parameters depend on yet to be determined the
effective cosmological constant represented through the AdS radius
$\ell$. The field equation for $\ell^{2}$ can be found by using
the effective parameters of (\ref{eq:eqa_eff_string}) in (\ref{eq:AdS_EoM_quad})
as 
\begin{equation}
\frac{1}{\ell^{2}}\left(1-\frac{\left(D-3\right)\left(D-4\right)\alpha^{\prime}}{4\ell^{2}}-\frac{\left(D-5\right)\left(D-6\right)\alpha^{\prime2}}{12\left(D-2\right)\ell^{4}}\right)=0.\label{eq:Cubic_string_AdS_EoM}
\end{equation}
Note that although we started with a quadratic equation in $1/\ell^{2}$,
that is (\ref{eq:AdS_EoM_quad}), we obtained a cubic equation as
expected for a cubic curvature theory. For $D\ge3$, there is always
an AdS solution as $1/\ell^{2}\sim\alpha^{\prime}$ in addition to
the flat solution, so that the theory admits an AdS-plane wave solution.

In addition to effective cosmological constant, we need the mass parameter
$M^{2}$ to write the AdS-plane wave solutions. Using EQCA parameters
in (\ref{eq:M2}), $M^{2}$ can be found as \cite{Gurses-PRL} 
\begin{equation}
M^{2}=\frac{4\ell^{2}}{7\alpha^{\prime2}}-\frac{2\left(D-3\right)\left(D-4\right)}{7\alpha^{\prime}}+\frac{29-9D}{7\ell^{2}}.\label{eq:Cubic_string_M2}
\end{equation}
The AdS-plane wave solutions given in the Introduction are the solutions
of (\ref{eq:Effective_string}) with this $M^{2}$. For example, for
the $\xi=0$ case one has 
\begin{equation}
V\left(u,z\right)=c_{1}\left(u\right)z^{D-3}+z^{\frac{D-5}{2}}\left(c_{2}\left(u\right)z^{\frac{1}{2}\sqrt{\left(D-1\right)^{2}+4\ell^{2}M^{2}}}+c_{3}\left(u\right)z^{-\frac{1}{2}\sqrt{\left(D-1\right)^{2}+4\ell^{2}M^{2}}}\right).
\end{equation}
Here, note that using the solutions of (\ref{eq:Cubic_string_AdS_EoM})
in (\ref{eq:Cubic_string_M2}), $M^{2}$ has the form $1/\alpha^{\prime}$
(as also dimensional analysis suggests) and becomes negative for $D>3$.
On the other hand, the BF bound, that is $M^{2}\ge-\frac{\left(D-1\right)^{2}}{4\ell^{2}}$,
is satisfied for $D\le6$.

To discuss pp-wave solutions, one should take the $\ell\rightarrow\infty$
limit in the AdS-plane wave field equations. Taking this limit in
(\ref{eq:Cubic_string_M2}) yields $M^{2}\rightarrow\infty$ which
suggests the absence of the massive operator part in the pp-wave field
equations. This is in fact the case which becomes more clear by taking
$\ell\rightarrow\infty$ limit at the EQCA level. In this limit, (\ref{eq:eqa_eff_string})
reduces to Einstein--Gauss-Bonnet theory which is the quadratic curvature
order of the original action (\ref{eq:Effective_string}).%
\footnote{This is expected since the EQCA is just the Taylor series expansion
in curvature around the maximally symmetric background. Then, once
the flat limit is taken in EQCA, the action reduces to the quadratic
curvature order of the original action.%
} Therefore, as we discussed above, the quadratic curvature order of
the original action determines the pp-wave field equations, and here
it is the Einstein--Gauss-Bonnet theory whose equations reduce to
the field equations of Einstein's gravity at the linearized level.
Therefore, the massive operator is absent and the pp-wave solutions
of (\ref{eq:Effective_string}) are only the Einsteinian solutions.

\subsection{Lanczos-Lovelock theory}

The Lanczos-Lovelock theory is a special $f\left(R_{\alpha\beta}^{\mu\nu}\right)$
theory which has at most second order derivatives of the metric in
its field equations just like Einstein's gravity. Therefore, one expects
a second order differential equation for the metric function $V$
as the (traceless) field equations for pp-waves and AdS-plane waves.
To find the explicit form of the field equations, one needs to construct
the EQCA for the Lanczos-Lovelock theory given with the Lagrangian
density 
\begin{equation}
f_{\text{L-L}}=\sum_{n=0}^{\left[\frac{D}{2}\right]}C_{n}\delta_{\nu_{1}\dots\nu_{2n}}^{\mu_{1}\dots\mu_{2n}}\prod_{p=1}^{n}R_{\mu_{2p-1}\mu_{2p}}^{\nu_{2p-1}\nu_{2p}},\label{eq:Lovelock}
\end{equation}
where $C_{n}$'s are dimensionful constants, $\delta_{\nu_{1}\dots\nu_{2n}}^{\mu_{1}\dots\mu_{2n}}$
is the generalized Kronecker delta, and $\left[\frac{D}{2}\right]$
denotes the integer part of its argument. In \cite{SpectraLovelock},
the EQCA of (\ref{eq:Lovelock}) was calculated as 
\begin{equation}
I_{\text{EQCA}}=\int d^{D}x\,\sqrt{-g}\left[\frac{1}{\tilde{\kappa}}\left(R-2\tilde{\Lambda}_{0}\right)+\tilde{\gamma}\left(R_{\alpha\beta}^{\mu\nu}R_{\mu\nu}^{\alpha\beta}-4R_{\nu}^{\mu}R_{\mu}^{\nu}+R^{2}\right)\right],
\end{equation}
where the effective parameters are: the effective Newton's constant,
\begin{equation}
\frac{1}{\tilde{\kappa}}\equiv2\left(D-2\right)!\sum_{n=0}^{\left[\frac{D}{2}\right]}\left(-1\right)^{n}C_{n}\frac{n\left(n-2\right)}{\left(D-2n\right)!}\left(\frac{2}{\ell^{2}}\right)^{n-1},\label{eq:k_tilde}
\end{equation}
the effective cosmological constant, 
\begin{equation}
\frac{\tilde{\Lambda}_{0}}{\tilde{\kappa}}\equiv-\frac{D!}{4}\sum_{n=0}^{\left[\frac{D}{2}\right]}\left(-1\right)^{n}C_{n}\frac{\left(n-1\right)\left(n-2\right)}{\left(D-2n\right)!}\left(\frac{2}{\ell^{2}}\right)^{n},\label{eq:L_tilde}
\end{equation}
and the effective Gauss-Bonnet coefficient 
\begin{equation}
\tilde{\gamma}\equiv2\left(D-4\right)!\sum_{n=0}^{\left[\frac{D}{2}\right]}\left(-1\right)^{n}C_{n}\frac{n\left(n-1\right)}{\left(D-2n\right)!}\left(\frac{2}{\ell^{2}}\right)^{n-2}.\label{eq:g_tilde}
\end{equation}
The AdS radius appearing in these effective parameters satisfies the
equation 
\begin{equation}
0=\sum_{n=0}^{\left[\frac{D}{2}\right]}\left(-2\right)^{n}C_{n}\frac{\left(D-2n\right)}{\left(D-2n\right)!}\left(\frac{1}{\ell^{2}}\right)^{n},\label{eq:LL_vacua}
\end{equation}
which can be found from (\ref{eq:AdS_EoM_quad}) after plugging (\ref{eq:k_tilde}-\ref{eq:g_tilde})
in. Again, notice that the quadratic equation (\ref{eq:AdS_EoM_quad})
yields an $\left[\frac{D}{2}\right]^{{\rm th}}$ order equation in
$1/\ell^{2}$ after the use of the $\ell$-dependent parameters of
EQCA. Note that for even dimensions, the $D=2n$ term does not contribute
to the field equation (\ref{eq:LL_vacua}). Since the EQCA is in the
Einstein--Gauss-Bonnet form, the traceless part of the field equations
reduces to 
\begin{equation}
\mathcal{O}V=0,\label{eq:LL-EoM}
\end{equation}
where $\mathcal{O}$ is defined in (\ref{eq:V_E_operators}), if $1/\tilde{\kappa}\ne0$,
that is 
\begin{equation}
\sum_{n=0}^{\left[\frac{D}{2}\right]}\left(-1\right)^{n}C_{n}\frac{n\left(D-2n\right)}{\left(D-2n\right)!}\left(\frac{2}{\ell^{2}}\right)^{n-1}\ne0,\label{eq:LL_constraint}
\end{equation}
holds. Thus, the Einsteinian solutions; for example, the Kaigorodov
solution 
\begin{equation}
V\left(u,z\right)=c\left(u\right)z^{D-3},
\end{equation}
solve (\ref{eq:LL-EoM}). Note that even though apparently $V_{E}$
does not show dependence on the parameters of the theory, the metric
depends on all the parameters via AdS radius $\ell$. Hence, the above
exercise is nontrivial.

Lastly, for the Chern-Simons Lovelock theory in odd dimensions \cite{CS-Lovelock},
the constraint $1/\tilde{\kappa}=0$ is satisfied, so the field equation
becomes trivial.

\section{Einsteinian Wave Solutions of the Generic Theory\label{sec:Einstein_soln_gen_theo}}

A natural generalization of the above exercises is that the AdS-plane
waves of the cosmological Einstein's theory solve the generic gravity
theory (\ref{eq:Generic_gravity}). The metric function $V$ does
not depend on the parameters of the theory; therefore, it is intact
for all theories. But, the nontrivial part of the computation is to
find the AdS radii for each theory. Fortunately, with the equivalent
linear action (ELA) procedure that we used in \cite{UnitaryBI,BI-Unitarity,AllUni3D},
all one needs to do is: 1- To calculate the Lagrangian density in
the maximally symmetric background (\ref{eq:Maximally_symmetric_background}),
let us call it $\bar{f}$; 2- To compute the derivative of the Lagrangian
density with respect to the Riemann tensor and evaluate it again in
(\ref{eq:Maximally_symmetric_background}), and with this result to
find 
\begin{align}
\left[\frac{\partial f}{\partial R_{\rho\sigma}^{\mu\nu}}\right]_{\bar{R}_{\rho\sigma}^{\mu\nu}}R_{\rho\sigma}^{\mu\nu} & \equiv\zeta R,
\end{align}
which is in fact definition of $\zeta$. Using these results, the
ELA, which has the same vacua as the original theory, can be constructed
as

\begin{equation}
I_{\text{ELA}}=\frac{1}{\tilde{\kappa}}\int d^{D}x\,\sqrt{-g}\left(R-2\tilde{\Lambda}_{0}\right),
\end{equation}
where the effective Newton's constant and the effective bare cosmological
constant are 
\begin{align}
\frac{1}{\tilde{\kappa}} & =\zeta,\\
\frac{\tilde{\Lambda}_{0}}{\tilde{\kappa}} & =-\frac{1}{2}\bar{f}-\frac{D\left(D-1\right)}{2\ell^{2}}\zeta.
\end{align}
Then, the AdS radii can be calculated from 
\begin{equation}
\ell^{2}=-\frac{\left(D-1\right)\left(D-2\right)}{2\tilde{\Lambda}_{0}}.
\end{equation}
Note that in the Lagrangian, the terms involving the derivatives of
the Riemann tensor do not contribute to the maximally symmetric vacua
at all. Because the field equations derived from these derivative
terms always involve the derivatives of the Riemann tensor which vanish
for the maximally symmetric metric.

As an example to this procedure, let us consider the conformal gravity
with derivative terms in $D=6$ dimensions.

\subsection*{Conformal gravity in $D=6$}

Conformal gravity in six dimensions%
\footnote{Note that to define a conformal gravity in six dimensions, one can
also use the two independent scalars constructed from three Weyl tensors,
see for example \cite{Pang,Tsoukalas}. For this purely cubic Weyl
theory, the EQCA and the linearized field equations will be identically
zero, so the AdS-plane wave field equations become trivial. The version
of six-dimensional conformal gravity we have chosen here is for discussing
the presence of derivatives of the Riemann tensor in the action.%
} has the Lagrangian density \cite{Henningson,Pang} 
\begin{equation}
\mathcal{L}_{{\rm Conf}}=\beta\left(RR_{\nu}^{\mu}R_{\mu}^{\nu}-\frac{3}{25}R^{3}-2R_{\nu}^{\mu}R_{\sigma}^{\rho}R_{\mu\rho}^{\nu\sigma}-R_{\nu}^{\mu}\square R_{\mu}^{\nu}+\frac{3}{10}R\square R\right).\label{eq:6D_conf_Lag}
\end{equation}
With the AdS-plane wave ansatz, the field equations coming from this
Lagrangian, which are given in \cite{Pang}, reduce to 
\begin{equation}
\left(\square+\frac{8}{\ell^{2}}\right)\left(\square+\frac{6}{\ell^{2}}\right)S_{\mu\nu}=0.\label{eq:Trless_eom_6D_conf}
\end{equation}
Using the ELA procedure above (see App.~\ref{sec:ELA_6D_conf}),
it is easy to show that for this purely cubic theory the AdS radius
is not fixed; therefore, any maximally symmetric space is a solution
which is to be expected since the theory is conformal with no internal
scale. But, once one imposes the existence of an AdS vacuum, one necessarily
breaks the symmetry in the vacuum and picks up a unique cosmological
constant (in \cite{Pang}, $\Lambda=-10$ was chosen in the $\ell=1$
units).

To further reduce (\ref{eq:Trless_eom_6D_conf}), using (\ref{eq:phi_id})
and (\ref{eq:Smn_in_hmn}) yields 
\begin{equation}
\left(\square+\frac{8}{\ell^{2}}\right)\left(\square+\frac{6}{\ell^{2}}\right)\left(\square+\frac{2}{\ell^{2}}\right)\left(\lambda_{\mu}\lambda_{\nu}V\right)=0,
\end{equation}
which still looks like a nonlinear differential equation since the
d'Alembertian operators are with respect to the full metric involving
the $V$ part. But, this apparent nonlinearity is a red-herring since
$\square^{n}\left(\lambda_{\alpha}\lambda_{\beta}V\right)=\bar{\square}^{n}\left(\lambda_{\alpha}\lambda_{\beta}V\right)$.
In addition, one can move $\lambda$ vectors to the left with the
help of (\ref{eq:phi_id}) and (\ref{eq:V_E_operators}) implying
\begin{equation}
\left(\square+\frac{2}{\ell^{2}}\right)\left(\lambda_{\alpha}\lambda_{\beta}V\right)=\left(\bar{\square}+\frac{2}{\ell^{2}}\right)\left(\lambda_{\alpha}\lambda_{\beta}V\right)=\lambda_{\alpha}\lambda_{\beta}\left(\frac{z^{2}}{\ell^{2}}\hat{\partial}^{2}-\frac{6}{\ell^{2}}\right)V,
\end{equation}
and get a linear differential equation 
\begin{equation}
\lambda_{\mu}\lambda_{\nu}\left(\frac{z^{2}}{\ell^{2}}\hat{\partial}^{2}\right)\left(\frac{z^{2}}{\ell^{2}}\hat{\partial}^{2}-\frac{2}{\ell^{2}}\right)\left(\frac{z^{2}}{\ell^{2}}\hat{\partial}^{2}-\frac{6}{\ell^{2}}\right)V=0.
\end{equation}
Assuming $V=V\left(u,z\right)$, the general solution reads 
\begin{equation}
V\left(u,z\right)=\frac{c_{1}}{z^{2}}+\frac{c_{2}}{z}+c_{3}+c_{4}z+c_{5}z^{2}+c_{6}z^{3},
\end{equation}
where the first term can be added to the ``background'' AdS part
and $c_{i}=c_{i}\left(u\right)$. Note that this is also the general
solution to the linearized equations with $h_{\mu\nu}=2V\lambda_{\mu}\lambda_{\nu}$.

To this conformal $D=6$ action, one can add the cosmological Einstein
and the Weyl square theory as \cite{Pang-BlackHole} 
\begin{equation}
\mathcal{L}_{6D}=R+\frac{20}{\ell^{2}}+\frac{\alpha}{2}C_{\alpha\beta}^{\mu\nu}C_{\mu\nu}^{\alpha\beta}-\mathcal{L}_{{\rm Conf}},
\end{equation}
whose field equations, upon the use of the AdS-plane wave ansatz,
reduce to 
\begin{equation}
\left[\beta\left(\square+\frac{8}{\ell^{2}}\right)\left(\square+\frac{6}{\ell^{2}}\right)+\frac{3}{2}\alpha\left(\square+\frac{6}{\ell^{2}}\right)+1\right]S_{\mu\nu}=0.\label{eq:6D_general_EoM}
\end{equation}
Unlike the purely cubic theory above, this theory has a unique vacuum
with $\Lambda=-10/\ell^{2}$ which is fixed by the cosmological Einstein
part: neither the quadratic Weyl piece nor the cubic part contributes
to the effective cosmological constant. Again assuming $V=V\left(u,z\right)$,
the general AdS-plane wave solution to (\ref{eq:6D_general_EoM})
with generic $\alpha$ and $\beta$ consists of six power terms 
\begin{equation}
V\left(u,z\right)=\sum_{i=1}^{6}c_{i}\left(u\right)z^{n_{i}},\label{eq:6D_gen_soln}
\end{equation}
with powers 
\begin{equation}
n_{1}=-2,\qquad n_{2}=3,\qquad n_{3,4,5,6}=\frac{1}{2}\pm\sqrt{\frac{5}{4}-\frac{3\alpha\ell^{2}}{4\beta}\pm\sqrt{\left(1+\frac{3\alpha\ell^{2}}{4\beta}\right)^{2}-\frac{\ell^{4}}{\beta}}},
\end{equation}
where again the first term can be added to the ``background'' AdS
part with no consequence. The second term is the Kaigorodov solution,
which can be expected without doing any calculation, and the rest
are the nontrivial pieces.

Let us consider the specific case of the ``tricritical gravity'',
that is $\alpha=-5\ell^{2}/12$ and $\beta=\ell^{4}/16$ \cite{Pang-BlackHole},
for which $n_{3,4,5,6}$ becomes $-2$ and $3$, so that the differential
equation (\ref{eq:6D_general_EoM}) degenerates into the form 
\begin{equation}
\lambda_{\mu}\lambda_{\nu}\left(\frac{z^{2}}{\ell^{2}}\hat{\partial}^{2}-\frac{6}{\ell^{2}}\right)^{3}V=0,
\end{equation}
with nontrivial logarithmic solutions in addition to the expected
Einsteinian parts, that are AdS and Kaigorodov parts, 
\begin{equation}
V\left(u,z\right)=\frac{1}{z^{2}}\left[c_{1}+c_{2}\ln\left(\frac{z}{\ell}\right)+c_{3}\ln^{2}\left(\frac{z}{\ell}\right)\right]+z^{3}\left[c_{4}+c_{5}\ln\left(\frac{z}{\ell}\right)+c_{6}\ln^{2}\left(\frac{z}{\ell}\right)\right],\label{eq:Log_soln_6D}
\end{equation}
where again $c_{i}=c_{i}\left(u\right)$. Note that both (\ref{eq:6D_gen_soln})
and (\ref{eq:Log_soln_6D}) are also the general solutions to the
corresponding linearized equations for transverse-traceless perturbations.

\section{Conclusion}

We have shown that the AdS-plane wave metric solves the most general
gravity theory whose Lagrangian is an arbitrary function of the metric,
the Riemann tensor and the covariant derivatives of the Riemann tensor.
In doing so, we have also given the explicit proof of the theorem,
briefly proved in \cite{Gurses-PRL}, that two-tensors in these spacetimes
can be written as a sum of $\square^{n}S_{\mu\nu}$ with $n=0,1,2,\dots$.
In our proof, the pp-wave solution played a role, so we have revisited
this spacetime and also constructed novel solutions for quadratic
gravity that also extend to the generic gravity theories. We have
devoted several sections to example theories such as the cubic curvature
gravity generated by string theory, Lanczos-Lovelock gravity, and
the recent $D=6$ conformal gravity, its nonconformal modifications,
and tricritical gravity.

Our exact solutions linearize the field equations, hence they also
match the perturbative solutions for transverse-traceless perturbations.
Therefore, the particle spectrum of these theories can be read from
these metrics, save the spin-0 mode. Once the unitarity constraints
on the particle spectrum are considered, this result can be used to
choose the viable theories. For example, requiring nontachyonic physical
excitations, that is $M_{n}^{2}=-b_{n}+\frac{2}{\ell^{2}}>0$ and
$\underset{\ell\rightarrow\infty}{\lim}M_{n}^{2}=M_{n,\text{flat}}^{2}>0$,
put tight constraints on the theory parameters such that they should
yield real, upper bounded $b_{n}$'s.

In obtaining our solutions, we have reduced the field equations of
the most general gravity theory to $N$ number of massive Klein-Gordon
equations satisfied by $V_{n}$, $n=1,2,\cdots,N$, and one massless
Klein Gordon equation satisfied by $V_{E}$ such that the general
solution to the theory is $V=V_{E}+\sum_{n=1}^{N}V_{n}$. We have
given sample solutions of these equations when all the masses are
different and when the masses are equal, that is the critical gravity
case. When some $M_{n}$'s are equal, the solutions involve logarithmic
parts. We gave the general logarithmic solution for the case where
all $M_{n}$'s are zero. The specific examples for this case are the
critical gravity theory studied in \cite{Alishah,Gullu-Gurses} and
the tricritical gravity theory that we discussed here.

We have generalized and unified the previous works \cite{Gueven,Horowitz-Steif,Coley}
on pp-wave spacetimes. Namely, for any theory the field equations
for the pp-wave metrics in the Kerr-Schild form reduce to 
\begin{equation}
\prod_{n=1}^{N}\left(\square+b_{n}\right)R_{\mu\nu}=0,
\end{equation}
which can be further reduced to the Einstein's gravity ones under
the assumptions of \cite{Gueven,Horowitz-Steif,Coley}. Another fact
is that the results we obtained in this work remain intact for the
theories with pure radiation sources, that is $T_{\mu\nu}dx^{\mu}dx^{\nu}=T_{uu}du^{2}$.
For example, for a source having the functional dependence $T_{\mu\nu}=T_{\mu\nu}\left(u\right)$,
the plane waves solving $a_{0}R_{\mu\nu}=T_{\mu\nu}$ are also particular
solutions to the generic gravity theory since $\square R_{\mu\nu}=0$
for these plane waves.

A possible setting to consider sources is introducing a nonminimally
coupled scalar field. In \cite{pp-waveScalar} and \cite{AdS-waveScalar},
it is shown that a nonminimally coupled scalar field with a specific
potential can support pp-wave and AdS-plane wave solutions for three-dimensional
Einstein's gravity. This specific potential form is generalized to
higher dimensions in \cite{pp-AdS_scalar_EoM}. Using this result
together with ones discussed here, it is possible to find pp-wave
and AdS-plane wave solutions to some higher curvature gravity theories
coupled to nonminimally coupled scalar fields \cite{CHOS}.

In a future work, the explicit proof presented for the AdS-plane waves
will be extended to Kerr-Schild--Kundt class discussed in \cite{gurses1,Gurses-PRL}.

\section{Acknowledgment}

M.~G. and B.~T. are supported by the TÜB\.{I}TAK grant 113F155.
M.~G. is a member of the Science Academy. T.~C.~S. is supported
by the Fondecyt grant 3140127. The Centro de Estudios Cient\'{i}ficos
(CECs) is funded by the Chilean Government through the Centers of
Excellence Base Financing Program of Conicyt. T.~C.~S. would like
to thank M.~Hassaine, E.~Ayon-Beato, and J.~Zanelli for useful
discussions.

\appendix

\section{Tensors of Kerr-Schild--Kundt Spacetimes\label{sec:KSK_tensors}}

In this section, we will prove (\ref{eq:Weyl_KSK}). For the Kerr-Schild--Kundt
class of metrics 
\begin{equation}
g_{\mu\nu}=\bar{g}_{\mu\nu}+2V\lambda_{\mu}\lambda_{\nu},\label{eq:AdS-wave_KS-1}
\end{equation}
where $\bar{g}_{\mu\nu}$ is the AdS metric and the following relations
hold 
\begin{equation}
\lambda^{\mu}\lambda_{\mu}=0,\qquad\nabla_{\mu}\lambda_{\nu}=\xi_{(\mu}\lambda_{\nu)},\qquad\xi_{\mu}\lambda^{\mu}=0,\qquad\lambda^{\mu}\partial_{\mu}V=0,\label{eq:AdS-wave_prop-1}
\end{equation}
the Riemann tensor is given in (B25) of \cite{gurses1} as 
\begin{align}
R_{\phantom{\mu}\alpha\nu\beta}^{\mu}= & \bar{R}_{\phantom{\mu}\alpha\nu\beta}^{\mu}+2\lambda_{\alpha}\lambda_{[\nu}\bar{\nabla}_{\beta]}\partial^{\mu}V-2\lambda^{\mu}\lambda_{[\nu}\bar{\nabla}_{\beta]}\partial_{\alpha}V\nonumber \\
 & +\lambda_{[\nu}\xi_{\beta]}\left(\lambda_{\alpha}\partial^{\mu}V-\lambda^{\mu}\partial_{\alpha}V+\lambda_{\alpha}\xi^{\mu}V\right)\nonumber \\
 & +\left(\lambda_{\alpha}\xi^{\mu}-\lambda^{\mu}\xi_{\alpha}\right)\lambda_{[\nu}\partial_{\beta]}V\nonumber \\
 & +2V\lambda^{\mu}\left(\lambda_{\alpha}\bar{\nabla}_{[\nu}\xi_{\beta]}-\lambda_{[\nu}\bar{\nabla}_{\beta]}\xi_{\alpha}\right).
\end{align}
One can free $\xi$ from derivatives with the help of $\left[\bar{\nabla}_{\mu},\bar{\nabla}_{\nu}\right]\lambda_{\beta}=\bar{R}_{\mu\nu\beta}^{\phantom{\mu\nu\beta}\rho}\lambda_{\rho}$
yielding 
\begin{equation}
2\lambda_{[\nu}\bar{\nabla}_{\mu]}\xi_{\beta}-2\lambda_{\beta}\bar{\nabla}_{[\nu}\xi_{\mu]}-\xi_{\beta}\lambda_{[\nu}\xi_{\mu]}=-\frac{2}{\ell^{2}}\left(g_{\mu\beta}\lambda_{\nu}-\lambda_{\mu}g_{\nu\beta}\right).
\end{equation}
Then, the Riemann tensor reduces to 
\begin{align}
R_{\phantom{\mu}\alpha\nu\beta}^{\mu}= & \bar{R}_{\phantom{\mu}\alpha\nu\beta}^{\mu}+\frac{4}{\ell^{2}}V\lambda^{\mu}\lambda_{[\nu}\bar{g}_{\beta]\alpha}\nonumber \\
 & +\lambda_{\alpha}\left(2\lambda_{[\nu}\bar{\nabla}_{\beta]}\partial^{\mu}V+\lambda_{[\nu}\xi_{\beta]}\partial^{\mu}V+\lambda_{[\nu}\xi_{\beta]}\xi^{\mu}V+\xi^{\mu}\lambda_{[\nu}\partial_{\beta]}V\right)\nonumber \\
 & -\lambda^{\mu}\left(2\lambda_{[\nu}\bar{\nabla}_{\beta]}\partial_{\alpha}V+\lambda_{[\nu}\xi_{\beta]}\partial_{\alpha}V+\lambda_{[\nu}\xi_{\beta]}\xi_{\alpha}V+\xi_{\alpha}\lambda_{[\nu}\partial_{\beta]}V\right),
\end{align}
whose $\left(0,4\right)$-rank tensor version is 
\begin{align}
R_{\mu\alpha\nu\beta}= & \bar{R}_{\mu\alpha\nu\beta}\nonumber \\
 & +\lambda_{\alpha}\left(2\lambda_{[\nu}\bar{\nabla}_{\beta]}\partial_{\mu}V+\lambda_{[\nu}\xi_{\beta]}\partial_{\mu}V+\lambda_{[\nu}\xi_{\beta]}\xi_{\mu}V+\xi_{\mu}\lambda_{[\nu}\partial_{\beta]}V\right)\nonumber \\
 & -\lambda_{\mu}\left(2\lambda_{[\nu}\bar{\nabla}_{\beta]}\partial_{\alpha}V+\lambda_{[\nu}\xi_{\beta]}\partial_{\alpha}V+\lambda_{[\nu}\xi_{\beta]}\xi_{\alpha}V+\xi_{\alpha}\lambda_{[\nu}\partial_{\beta]}V\right).
\end{align}

Now, let us calculate the Weyl tensor 
\begin{equation}
C_{\mu\alpha\nu\beta}\equiv R_{\mu\alpha\nu\beta}-\frac{2}{D-2}\left(g_{\mu[\nu}R_{\beta]\alpha}-g_{\alpha[\nu}R_{\beta]\mu}\right)+\frac{2}{\left(D-1\right)\left(D-2\right)}Rg_{\mu[\nu}g_{\beta]\alpha}.
\end{equation}
The last term involving the scalar curvature has the form 
\begin{equation}
Rg_{\mu[\nu}g_{\beta]\alpha}=\bar{R}\bar{g}_{\mu[\nu}\bar{g}_{\beta]\alpha}-2\bar{R}V\left(\lambda_{\alpha}\lambda_{[\nu}\bar{g}_{\beta]\mu}-\lambda_{\mu}\lambda_{[\nu}\bar{g}_{\beta]\alpha}\right).
\end{equation}
The second term involving the Ricci tensor has the following form
by using $R_{\alpha\beta}=-\frac{\left(D-1\right)}{\ell^{2}}g_{\alpha\beta}+\rho\lambda_{\alpha}\lambda_{\beta}$,
\begin{equation}
g_{\mu[\nu}R_{\beta]\alpha}-g_{\alpha[\nu}R_{\beta]\mu}=\bar{g}_{\mu[\nu}\bar{R}_{\beta]\alpha}-\bar{g}_{\alpha[\nu}\bar{R}_{\beta]\mu}-\left(\rho+\frac{4V\bar{R}}{D}\right)\left(\lambda_{\alpha}\lambda_{[\nu}\bar{g}_{\beta]\mu}-\lambda_{\mu}\lambda_{[\nu}\bar{g}_{\beta]\alpha}\right).
\end{equation}
With the help of the above results and $\bar{C}_{\mu\alpha\nu\beta}=0$,
the Weyl tensor reduces to 
\begin{align}
C_{\mu\alpha\nu\beta}= & \lambda_{\alpha}\left(2\lambda_{[\nu}\bar{\nabla}_{\beta]}\partial_{\mu}V+\lambda_{[\nu}\xi_{\beta]}\partial_{\mu}V+\lambda_{[\nu}\xi_{\beta]}\xi_{\mu}V+\xi_{\mu}\lambda_{[\nu}\partial_{\beta]}V\right)\nonumber \\
 & -\lambda_{\mu}\left(2\lambda_{[\nu}\bar{\nabla}_{\beta]}\partial_{\alpha}V+\lambda_{[\nu}\xi_{\beta]}\partial_{\alpha}V+\lambda_{[\nu}\xi_{\beta]}\xi_{\alpha}V+\xi_{\alpha}\lambda_{[\nu}\partial_{\beta]}V\right)\nonumber \\
 & +\frac{2}{D-2}\left(\rho-\frac{2\left(D-2\right)}{\ell^{2}}V\right)\left(\lambda_{\alpha}\lambda_{[\nu}\bar{g}_{\beta]\mu}-\lambda_{\mu}\lambda_{[\nu}\bar{g}_{\beta]\alpha}\right),
\end{align}
where one can convert $\bar{g}_{\mu\nu}$ to $g_{\mu\nu}$ without
producing any additional term. Also, using (B4) and (B5) of \cite{gurses1},
one has 
\begin{equation}
\bar{\nabla}_{\alpha}\partial_{\beta}V=\nabla_{\alpha}\partial_{\beta}V-\lambda_{\alpha}\lambda_{\beta}\left(\partial^{\lambda}V\right)\partial_{\lambda}V,
\end{equation}
which can be used to write $C_{\mu\alpha\nu\beta}$ in terms of the
full metric quantities as 
\begin{align}
C_{\mu\alpha\nu\beta}= & \lambda_{\mu}\lambda_{\nu}\left[-\nabla_{\alpha}\partial_{\beta}V-\xi_{(\alpha}\partial_{\beta)}V-\frac{1}{2}\xi_{\alpha}\xi_{\beta}V-\frac{1}{D-2}g_{\alpha\beta}\left(\rho-\frac{2\left(D-2\right)}{\ell^{2}}V\right)\right]\nonumber \\
 & +\lambda_{\alpha}\lambda_{\beta}\left[-\nabla_{\mu}\partial_{\nu}V-\xi_{(\mu}\partial_{\nu)}V-\frac{1}{2}\xi_{\mu}\xi_{\nu}V-\frac{1}{D-2}g_{\mu\nu}\left(\rho-\frac{2\left(D-2\right)}{\ell^{2}}V\right)\right]\nonumber \\
 & -\lambda_{\mu}\lambda_{\beta}\left[-\nabla_{\alpha}\partial_{\nu}V-\xi_{(\alpha}\partial_{\nu)}V-\frac{1}{2}\xi_{\alpha}\xi_{\nu}V-\frac{1}{D-2}g_{\alpha\nu}\left(\rho-\frac{2\left(D-2\right)}{\ell^{2}}V\right)\right]\nonumber \\
 & -\lambda_{\alpha}\lambda_{\nu}\left[-\nabla_{\mu}\partial_{\beta}V-\xi_{(\mu}\partial_{\beta)}V-\frac{1}{2}\xi_{\mu}\xi_{\beta}V-\frac{1}{D-2}g_{\mu\beta}\left(\rho-\frac{2\left(D-2\right)}{\ell^{2}}V\right)\right].
\end{align}
Thus, defining 
\begin{equation}
\Omega_{\alpha\beta}=-\left[\nabla_{\alpha}\partial_{\beta}V+\xi_{(\alpha}\partial_{\beta)}V+\frac{1}{2}\xi_{\alpha}\xi_{\beta}V+\frac{1}{D-2}g_{\alpha\beta}\left(\rho-\frac{2\left(D-2\right)}{\ell^{2}}V\right)\right],
\end{equation}
reduces $C_{\mu\alpha\nu\beta}$ in the desired Type-N form 
\begin{align}
C_{\mu\alpha\nu\beta} & =\lambda_{\mu}\lambda_{\nu}\Omega_{\alpha\beta}+\lambda_{\alpha}\lambda_{\beta}\Omega_{\mu\nu}-\lambda_{\mu}\lambda_{\beta}\Omega_{\alpha\nu}-\lambda_{\alpha}\lambda_{\nu}\Omega_{\mu\beta}\nonumber \\
 & =4\lambda_{[\mu}\Omega_{\alpha][\beta}\lambda_{\nu]}.
\end{align}

\section{Bianchi Identities For The Weyl Tensor\label{sec:Bianchi_Weyl}}

Once-contracted Bianchi identity 
\begin{equation}
\nabla^{\nu}R_{\mu\alpha\nu\beta}=\nabla_{\mu}R_{\alpha\beta}-\nabla_{\alpha}R_{\mu\beta},
\end{equation}
for constant $R$ yields 
\begin{equation}
\nabla^{\nu}R_{\mu\alpha\nu\beta}=\nabla_{\mu}S_{\alpha\beta}-\nabla_{\alpha}S_{\mu\beta},
\end{equation}
which then leads to

\begin{equation}
\nabla^{\mu}C_{\mu\alpha\nu\beta}=\nabla_{\nu}S_{\beta\alpha}-\nabla_{\beta}S_{\nu\alpha}-\frac{2}{D-2}\nabla^{\mu}\left(g_{\mu[\nu}S_{\beta]\alpha}-g_{\alpha[\nu}S_{\beta]\mu}\right).
\end{equation}
Using the twice-contracted Bianchi identity, that is $\nabla^{\mu}S_{\mu\nu}=0$
for constant $R$, one gets 
\begin{equation}
\nabla^{\mu}C_{\mu\alpha\nu\beta}=\frac{D-3}{D-2}\left(\nabla_{\nu}S_{\beta\alpha}-\nabla_{\beta}S_{\nu\alpha}\right),
\end{equation}
which is the once-contracted Bianchi identity of the Weyl tensor for
constant curvature spacetimes.

Now, let us discuss $\nabla^{\mu}\nabla^{\nu}C_{\mu\alpha\nu\beta}$
which becomes 
\begin{equation}
\nabla^{\mu}\nabla^{\nu}C_{\mu\alpha\nu\beta}=\frac{D-3}{D-2}\left(\square S_{\alpha\beta}-\nabla^{\mu}\nabla_{\alpha}S_{\mu\beta}\right),
\end{equation}
for constant curvature spacetimes. Then, using $\nabla^{\mu}\nabla_{\sigma}S_{\mu\nu}=\frac{R}{D-1}S_{\sigma\nu}$,
which holds for the metrics (\ref{eq:AdS-wave_KS-1}), one gets 
\begin{equation}
\nabla^{\mu}\nabla^{\nu}C_{\mu\alpha\nu\beta}=\frac{D-3}{D-2}\left(\square S_{\alpha\beta}-\frac{R}{D-1}S_{\alpha\beta}\right),
\end{equation}
which proves (\ref{eq:Bianchi_Weyl}).

\section{Equivalent Linear Action of Conformal Gravity\label{sec:ELA_6D_conf}}

Without finding the complicated field equations of the six-dimensional
conformal gravity, let us show a method that leads to the effective
cosmological constant. First, note that the effective cosmological
constant of a generic gravity theory is determined by only the nonderivative
Riemann terms appearing in the action. Because the field equations
derived from the terms involving the derivative of the Riemann tensor
always yield zero after evaluating them for the maximally symmetric
background 
\begin{equation}
\bar{R}_{\rho\sigma}^{\mu\nu}=-\frac{1}{\ell^{2}}\left(\delta_{\rho}^{\mu}\delta_{\sigma}^{\nu}-\delta_{\sigma}^{\mu}\delta_{\rho}^{\nu}\right).
\end{equation}
Thus, we need to focus on the terms involving the Riemann tensor but
not its derivatives. The procedure described in Sec.~\ref{sec:Einstein_soln_gen_theo}
for construction of the ELA is based on the following Taylor series
expansion of the Lagrangian density 
\begin{align}
f_{\text{ELA}}\left(R_{\alpha\beta}^{\mu\nu}\right)= & f\left(\bar{R}_{\alpha\beta}^{\mu\nu}\right)+\left[\frac{\partial f}{\partial R_{\eta\theta}^{\rho\sigma}}\right]_{\bar{R}_{\eta\theta}^{\rho\sigma}}\left(R_{\eta\theta}^{\rho\sigma}-\bar{R}_{\eta\theta}^{\rho\sigma}\right).
\end{align}
However, in doing computations, one may prefer to consider the functional
dependence of the $f\left(R_{\alpha\beta}^{\mu\nu}\right)$ theory
as $f\left(R,R_{\nu}^{\mu},R_{\alpha\beta}^{\mu\nu}\right)$ and one
has 
\begin{align}
f_{\text{ELA}}\left(R,R_{\nu}^{\mu},R_{\alpha\beta}^{\mu\nu}\right)= & f\left(\bar{R},\bar{R}_{\nu}^{\mu},\bar{R}_{\alpha\beta}^{\mu\nu}\right)+\left[\frac{\partial f}{\partial R}\right]_{\bar{R}}\left(R-\bar{R}\right)+\left[\frac{\partial f}{\partial R_{\sigma}^{\rho}}\right]_{\bar{R}_{\sigma}^{\rho}}\left(R_{\sigma}^{\rho}-\bar{R}_{\sigma}^{\rho}\right)\nonumber \\
 & +\left[\frac{\partial f}{\partial R_{\eta\theta}^{\rho\sigma}}\right]_{\bar{R}_{\eta\theta}^{\rho\sigma}}\left(R_{\eta\theta}^{\rho\sigma}-\bar{R}_{\eta\theta}^{\rho\sigma}\right).
\end{align}
Using this formula, let us construct the ELA for each nonderivative
Riemann term in the (\ref{eq:6D_conf_Lag}). First, note that the
background Ricci tensor and the background scalar curvature in six
dimensions in our conventions are 
\begin{equation}
\bar{R}=-\frac{30}{\ell^{2}},\qquad\bar{R}_{\nu}^{\mu}=-\frac{5}{\ell^{2}}\delta_{\nu}^{\mu}.
\end{equation}
Then, for the term $f\left(R,R_{\nu}^{\mu}\right)=RR_{\nu}^{\mu}R_{\mu}^{\nu}$,
one needs $\bar{f}$ and $\zeta$ to construct ELA which are 
\begin{equation}
\bar{R}\bar{R}_{\nu}^{\mu}\bar{R}_{\mu}^{\nu}=-\frac{4500}{\ell^{6}},\qquad\zeta=\frac{450}{\ell^{4}},
\end{equation}
and the ELA for $f\left(R,R_{\nu}^{\mu}\right)=RR_{\nu}^{\mu}R_{\mu}^{\nu}$
becomes 
\begin{equation}
f_{\text{ELA}}\left(R,R_{\nu}^{\mu}\right)=\frac{450}{\ell^{4}}\left(R+\frac{20}{\ell^{2}}\right).
\end{equation}
Moving the $R^{3}$ term, $\bar{f}$ and $\zeta$ become 
\begin{equation}
\bar{R}^{3}=-\frac{27000}{\ell^{6}},\qquad\zeta=\frac{2700}{\ell^{4}},
\end{equation}
which yields the ELA, 
\begin{equation}
f_{\text{ELA}}\left(R\right)=\frac{2700}{\ell^{4}}\left(R+\frac{20}{\ell^{2}}\right).
\end{equation}
Finally, for the term $f\left(R_{\nu}^{\mu},R_{\alpha\beta}^{\mu\nu}\right)=R_{\nu}^{\mu}R_{\sigma}^{\rho}R_{\mu\rho}^{\nu\sigma}$,
$\bar{f}$ and $\zeta$ can be calculated as 
\begin{equation}
\bar{f}=-\frac{750}{\ell^{6}},\qquad\zeta=\frac{75}{\ell^{4}},
\end{equation}
and the ELA becomes 
\begin{equation}
f_{\text{ELA}}\left(R_{\nu}^{\mu},R_{\alpha\beta}^{\mu\nu}\right)=\frac{75}{\ell^{4}}\left(R+\frac{20}{\ell^{2}}\right).
\end{equation}
Collecting all these results yields the ELA for (\ref{eq:6D_conf_Lag})
as 
\begin{align*}
f_{\text{ELA}}^{\text{6D-conf}}= & -\frac{24}{\ell^{4}}\left(R+\frac{20}{\ell^{2}}\right),
\end{align*}
whose vacuum equation is 
\[
\ell^{2}=-\frac{\left(D-1\right)\left(D-2\right)}{2\tilde{\Lambda}_{0}}=\ell^{2}.
\]
Thus, AdS with any cosmological constant is a solution as expected
in this scale-free theory.

\end{document}